\documentclass[12pt]{article}

\usepackage[utf8]{inputenc}
\usepackage{amsmath,amsfonts,graphicx}
\usepackage[square,numbers,sort&compress]{natbib}
\usepackage{fullpage}
\usepackage{lineno}
\usepackage{xcolor}
\usepackage{pdflscape}
\usepackage{hyperref}
\usepackage{tcolorbox}
\usepackage{authblk}

\tcbuselibrary{breakable}

\newcommand{\vect}[1]{\boldsymbol{\mathbf{#1}}}

\newcommand{\highlight}{}                   

\title{Random walk models in the life sciences: including births, deaths and local interactions}

\author[1,*]{Michael J. Plank}
\author[2,3]{Matthew J. Simpson}
\author[4]{Ruth E. Baker}

\affil[1]{School of Mathematics and Statistics, University of Canterbury, Christchurch, New Zealand}
\affil[2]{School of Mathematical Sciences, Queensland University of Technology, Brisbane, Australia}
\affil[3]{ARC Centre of Excellence for the Mathematical Analysis of Cellular Systems, QUT,  Brisbane,  Australia.}
\affil[4]{Mathematical Institute, University of Oxford, Oxford, UK}

\affil[*]{Corresponding author: Michael J. Plank, michael.plank@canterbury.ac.nz}

\date{}

\begin{document}


\maketitle


\begin{abstract}
Random walks and related spatial stochastic models have been used in a range of application areas including animal and plant ecology, infectious disease epidemiology, developmental biology, wound healing, and oncology. Classical random walk models assume that all individuals in a population behave independently, ignoring local physical and biological interactions. This assumption simplifies the mathematical description of the population considerably, enabling continuum-limit descriptions to be derived and used in model analysis and fitting. However, interactions between individuals can have a crucial impact on population-level behaviour. In recent decades, research has increasingly been directed towards models that include interactions, including physical crowding effects and local biological processes such as adhesion, competition, dispersal, predation and adaptive directional bias. In this article, we review the progress that has been made with models of interacting individuals. We aim to provide an overview that is accessible to researchers in application areas, as well as to specialist modellers. We focus particularly on derivation of asymptotically exact or approximate continuum-limit descriptions and simplified deterministic models of mean-field behaviour and resulting spatial patterns. We provide worked examples and illustrative results of selected models. We conclude with a discussion of current areas of focus and future challenges. 
\end{abstract}



\clearpage

\section{Introduction}


Mathematical modelling is playing an increasingly pivotal role in many areas of the life sciences. Models can make a range of contributions including: providing insight into the biological mechanisms underlying observed patterns in empirical data; acting as surrogates for experiments that would be too costly or unethical to conduct in practice; generating hypotheses that can be tested experimentally; interpreting and predicting empirical data; quantifying uncertainty in biological parameter estimates; and comparing the effects of alternative interventions and control strategies, such as medical treatments or biodiversity conservation programmes. 

There is a broad class of models that represent the distribution of a population of agents in space and time. The agents may represent cells, plants, animals or humans depending on the application. The state of the population (i.e.~the number of agents and their locations and other attributes) changes over time via processes such as migration, proliferation, dispersal, and death. The population can be described by a stochastic agent-based model (ABM), sometimes also known as an individual-based model, which explicitly tracks individual agents over space and time. Alternatively, the population may be described by a deterministic continuum-limit model, which typically represents the mean density of agents as a continuous function of space and time. ABMs allow individual-level mechanisms and characteristics to be easily specified and simulated. In contrast, deterministic continuum-limit models are usually more computationally efficient and can offer greater mechanistic insight into the effect of different parameter values and dynamic regimes, such as equilibrium or asymptotic behaviour and bifurcations. There are a variety of approaches for deriving exact or approximate continuum-limit equations from an ABM.

In many biological application areas, modelling interactions among individual agents is important to accurately capture and provide insight into biological mechanisms and the role of individual variability and fluctuations. One key area of application is ecology, and in particular plant ecology where the growth or death of individuals is influenced by local competition (e.g. for space, nutrients or sunlight) and seed dispersal~\cite{kot1996dispersal,bolker1997using}. These kinds of interactions governing birth-death processes are important in other applications, such as modelling the dynamics of populations of cells. A major difference between plant ecology and cell biology is that biological cells are typically very motile, so interactions such as crowding effects~\cite{simpson2007simulating} and local attraction and repulsion cues~\cite{Schmeiser2009} are important to account for in models. Models used in animal movement ecology also need to account for interactions affecting movement, such as group navigation~\cite{codling2007group} and predator-prey interactions \cite{murrell2005local}. 

The mathematical theory for populations of non-interacting agents independently undergoing random or directed motility has been well studied~\cite{berg1983random,okubo1989spatial,codling2008random}. When agents behave independently, the system is linear and population-level quantities can be obtained simply by averaging over independent realisations of a single-agent model. In contrast, when agents interact with one another, this typically makes the system nonlinear, meaning that the whole is no longer just the sum of the parts. In the terminology of Weaver~\cite{weaver1948science}, interactions among agents transform the system from being a problem of disorganised complexity, amenable to the techniques of statistical physics, to a problem of organised complexity. Understanding the behaviour of such systems is a more recent and ongoing area of active research.

The derivation of deterministic continuum-limit equations that asymptotically or approximately describe how the average population density, or other quantities of interest, depend on time and/or space is a key focus. Stochastic ABMs can be simulated computationally. However, there are several advantages to having even an approximate continuum-limit model for key quantities of interest in the ABM. Firstly, ABMs are often computationally expensive and typically require a large number of realisations to produce reliable statistics. Secondly, continuum-limit equations can reveal how microscopic individual-level mechanisms affect (and in some case are affected by) macroscopic, population-level quantities. For example, how do factors influencing the proliferation and death of individuals affect the total population size? How are individual trajectories linked to the changing spatial distribution of the population over time? How does micro-scale individual behaviour interact with macro-scale spatial structure, such as the tendency of individuals to cluster together in space? Thirdly, continuum-limit equations can deliver insights into the key parameter dependencies in the model, which tend to be opaque to simulation-only approaches. Fourthly, computationally efficient deterministic approximations are typically more amenable to methods for fitting models to empirical data, parameter inference and uncertainty quantification. 


\subsection*{Aims and structure of this review}

In this review, we briefly summarise the classical theory of random walks of non-interacting agents and then review some of the literature on models that include interactions of different types. We aim to provide an overview that is accessible to early-career researchers and to researchers in relevant application areas in the life sciences. We therefore omit some of the technical details, but provide references to the mathematical literature for the interested reader.   Throughout, we point out the mathematical connections between the various modelling approaches and their links to application areas. We do not attempt to give an exhaustive literature review,  but instead focus on some of the key advances and selected examples that illustrate important concepts in the area.

In Section~\ref{sec:non_interacting}, we review the classical theory of random walks of non-interacting individuals and their associated continuum-limit equations. In Section~\ref{sec:crowding}, we show how the models and techniques for deriving the continuum limit introduced in Section~\ref{sec:non_interacting} can be generalised to include crowding effects, where individuals compete for space. As we will see, including interactions between individuals gives rise, in some cases, to nonlinear terms in the continuum-limit equation. In Section~\ref{sec:other_interactions}, we cover other types of interactions between individuals, such as adhesion or repulsion, and some model extensions. In Section~\ref{sec:spatial_structure}, we show how individual behaviour affects and is affected by the spatial structure of the population. This enables models to go beyond the spatial mean-field equations introduced in Sections \ref{sec:crowding}--\ref{sec:other_interactions} by accounting for spatial correlations between the locations of individuals within the population.

In Boxes 1--5, we provide optional worked examples illustrating the fundamental concepts behind the ABMs and the techniques used to derive population-level continuum descriptions. We also provide example results from stochastic ABM simulations and numerical solutions of corresponding population-level models. Code to reproduce these results is available at \url{https://github.com/michaelplanknz/interacting-random-walk-models}. A list of mathematical notation used throughout the paper is provided in Supplementary Table S1.
 

\section{Random walks of non-interacting individuals}
\label{sec:non_interacting}

In this section, we begin by reviewing some of the fundamental theory of random walks and their continuum-limit equations. This theory, which dates back to the work of Pearson~\cite{pearson2005problem} and Rayleigh~\cite{rayleigh2005problem} in 1905, applies to collections of individuals that are assumed to be moving independently of one another, and it plays a prominent role in statistical mechanics~\cite{casquilho2015introduction}. We show how random and directed individual-level motility mechanisms give rise to diffusive and advective processes, respectively, at the macroscopic scale. We also briefly highlight some of the differences between lattice-based and lattice-free models, and position-jump and velocity-jump processes. For generality, we present the theory in $d$ spatial dimensions. However, most applications and examples will be for the two-dimensional case. This preliminary section lays the groundwork for the subsequent sections, which will cover models of interacting individuals. In this more complicated situation, the movement, proliferation and/or death rates for one agent can depend on the number and location of other agents in the population.


\subsection{Random walk fundamentals}
\label{sec:fundamentals}

A {\em position-jump random walk} is a type of stochastic process that is a discrete-time Markov chain for the location $\vect{X}_n$ of a random walker, known as an {\em agent}, after $n$ steps. In the case of an {\em unrestricted} random walk on an unbounded domain $\Omega\subseteq\mathbb{R}^d$, the process can be defined by
\begin{equation} 
\label{eq:rw_defn}
\vect{X}_0 = \vect{x}_0, \quad \vect{X}_{n}=\vect{X}_{n-1}+\vect{Z}_n \quad \textrm{ for } n=1,2,\ldots,
\end{equation}
where the $\vect{Z}_n$ are independent, identically distributed (IID) random variables, sometimes called {\em jumps}. We typically assume that each random walk step takes fixed duration $\tau$, and we sometimes denote the agent's location by $\vect{X}(t)$ instead of $\vect{X}_n$ with $t=n\tau$.

The random walk can be {\em restricted} to a bounded domain $\Omega\subset \mathbb{R}^d$ via boundary conditions that specify what happens in cases where $\vect{X}_{n-1}+\vect{Z}_n\notin \Omega$. Common examples include: absorbing boundaries, where if the agent reaches the boundary of $\Omega$ it remains there for all future time; reflecting boundaries, where the agent is reflected back into the interior of $\Omega$; and periodic boundaries, where the agent's location is wrapped around to the opposite side of the domain. 


\subsection{Lattice-based random walks} 
\label{sec:non_interacting_lattice_based}

An important subcategory of the process defined by Equation~\eqref{eq:rw_defn} is {\em lattice-based random walks}, in which the agent's location is restricted to a regular lattice of spacing $\delta$. The most common choice is a square lattice, although in two dimensions a hexagonal lattice can be used \cite{aubert2006cellular,fernando2010nonlinear} and approaches have also been developed for unstructured meshes~\cite{Engblom:2009:SSR,Lotstedt:2015:SOS}. On a square lattice, the jumps must satisfy $\vect{Z}_n/\delta \in \mathbb{Z}^d$. All examples of lattice-based random walks considered here will be {\em nearest-neighbour} walks on a square lattice. This means that the agent can only move to an adjacent lattice site, and so the jump distribution has support on $\pm\delta \hat{\vect{e}}_k$, where $\hat{\vect{e}}_k$ is the unit vector in the $k^\text{th}$ Cartesian coordinate direction.


\subsubsection{Unbiased lattice-based random walks}

In an {\em unbiased random walk}, the expected value of the jump variable, $\langle \vect{Z}_n\rangle$, is zero so there is no change in the expected agent location from one step to the next. The simplest example is a nearest-neighbour lattice-based random walk where the agent has fixed probability $P_m$ of moving during each time step of duration $\tau$ and equal probabilities of moving to each adjacent lattice site. For example, in the two-dimensional case, this implies that $\vect{Z}_n$ takes values $(\delta,0)$,  $(-\delta,0)$, $(0,\delta)$, or $(0,-\delta)$ with probability $P_m/4$ each, and value $(0,0$) with probability $1-P_m$. As the time step $\tau$ and lattice spacing $\delta$ tend to zero, the probability density function $p_1(\vect{x},t)$ for the location $\vect{X}(t)$ of the agent at time $t=n\tau$ satisfies the {\em linear diffusion equation} (also known as the Fickian diffusion equation or the heat equation)~\cite{lin1974mathematics}
\begin{equation} 
\label{eq:heat}
\frac{\partial p_1}{\partial t} = D\nabla^2 p_1,
\end{equation}
where $D>0$ is the {\em diffusivity} (sometimes referred to as the diffusion coefficient) given by $D=\lim_{\delta,\tau\to 0} P_m\delta^2/(2d\tau)$, see~\cite{codling2008random} for more details. For this limit to exist and be non-zero requires the ratio $\delta^2/\tau$ to be held constant as $\delta$ and $\tau$ jointly tend to zero.  Equation~\eqref{eq:heat} is known as the {\em continuum-limit} equation for the random walk, because in the limiting case $\delta,\tau\to 0$ the variables $\vect{x}$, $t$ and $p_1$ are treated as continuous as opposed to discrete variables. Note that, for a given time $t=n\tau$, taking the limit $\tau\to 0$ implies that $n\to\infty$, and hence the continuum-limit should be viewed as a valid approximation after a sufficiently large number of random walk steps (and at a length-scale that is large relative to the step size $\delta$). 

In Equation~\eqref{eq:heat} and most of the continuum-limit equations introduced later in this article, the location variable $\vect{x}$ may belong to the unbounded $d$-dimensional domain $\mathbb{R}^d$ or to some bounded domain $R\subset \mathbb{R}^d$, depending on whether the random walk is unrestricted or restricted in space. Suitable boundary conditions on $p_1$ are needed and these can include absorbing, reflecting (i.e.~no-flux), or periodic boundary conditions. The time variable $t$ satisfies $t\ge0$, with an initial condition required for $p_1(\vect{x},0)$. 

The simplest interpretation of $p_1(\vect{x},t)$ in Equation~\eqref{eq:heat} is as the probability density function for the location of a single agent at time $t$. However, if there are $N$ agents undergoing {\em independent} random walks then, because Equation~\eqref{eq:heat} is linear, the sum of their probability density functions $p(\vect{x},t)=\sum_{k=1}^N p_k(\vect{x},t)$ also satisfies Equation~\eqref{eq:heat}. This requires that the movement probabilities of agent $k$ do not depend on the locations of other agents. In this case, $p(\vect{x},t)$ can be interpreted as the {\em expected agent density} (i.e.~mean number of agents per unit volume, with dimensions $(\mathrm{length}
)^{-d}$) at location $\vect{x}$ and time $t$, such that $\int_R p(\vect{x},t)\,\text{d}\vect{x} = N$ for all values of $t$. From here on, we will mainly work with the expected agent density $p(\vect{x},t)$ in deriving continuum-limit equations, reflecting our focus on populations of agents. 

\begin{figure}
\centering
	\includegraphics[width=\textwidth]{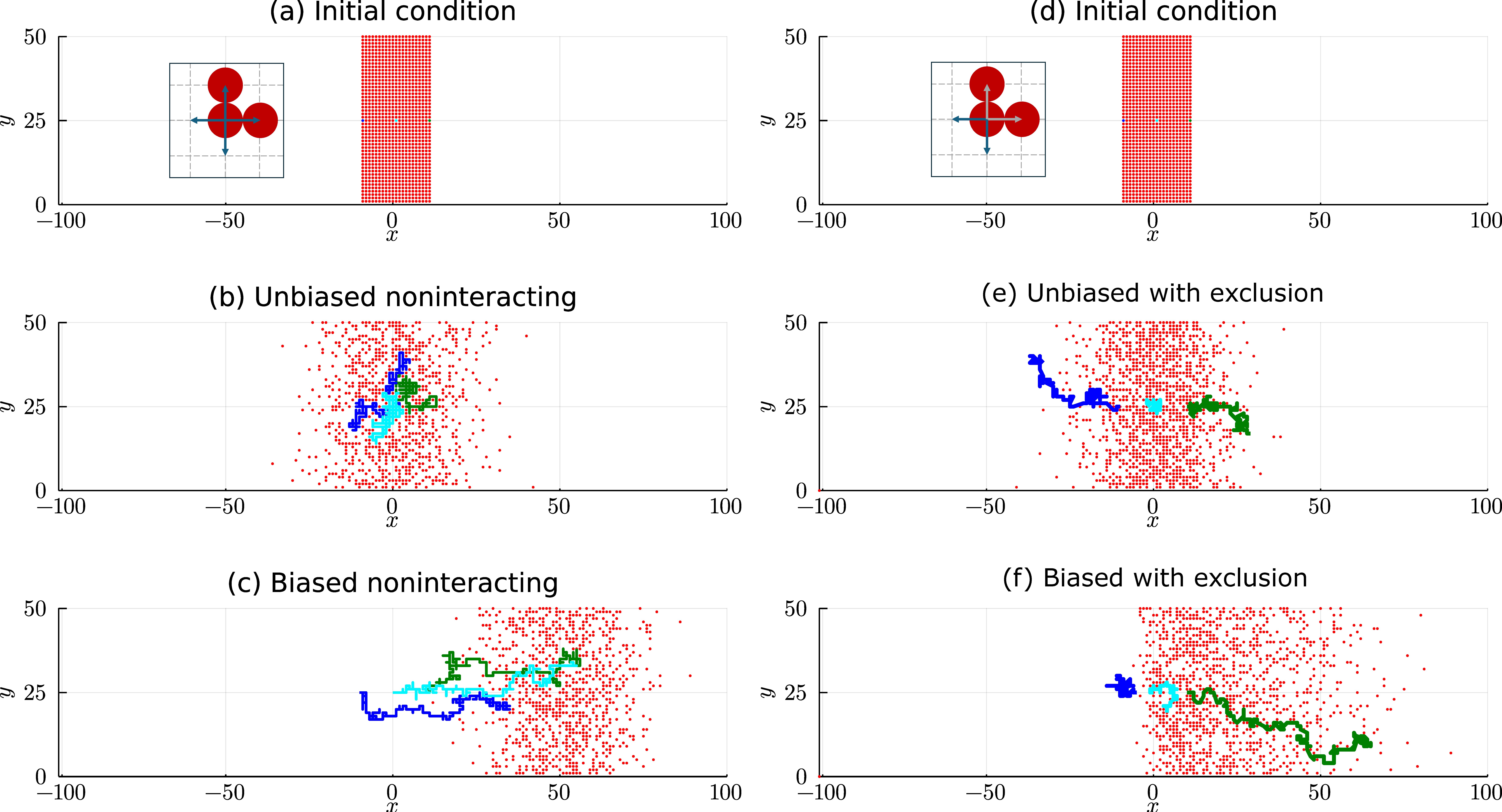}
	\caption{ {\highlight Comparison of lattice-based random walks models with and without interactions between agents. Results in the left column correspond to a non-interacting random walk (see inset in panel a -- movement in each of the four orthogonal directions is always allowed regardless of other agents); results in the right column correspond to an interacting exclusion process (see inset in panel b -- attempted moves onto lattice sites that are already occupied are aborted as indicated by grey arrows).} Simulations are on a regular lattice of dimensions $201 \times 51$ with lattice spacing $\delta = 1$ such that $-100 \le x \le 100$ and $0 \le y \le 50$, as indicated. The initial conditions (a,d) are identical and consist of a population in which one agent is placed at each lattice site with $-10 \le x \le 10$ and all other lattice sites are empty. Three agents initially at locations $(-10,25)$, $(0,25)$ and $(10,25)$, coloured blue, cyan and green, respectively, are tagged and their trajectories shown. All remaining agents are coloured red and only the final positions of the red agents are shown, not their trajectories. Snapshots in the middle row (b,e) show the result of performing an unbiased random walk with $P_m=1$ and $\rho_x=0$ over 200 time steps. Simulation snapshots in the bottom row (c,f) show the result of performing a biased random walk with $P_m=1$ and $\rho_x=1/2$ over 200 time steps.}
	\label{fig:scatter_plots}
\end{figure}

To make these ideas more concrete we present some stochastic simulation results in Figure~\ref{fig:scatter_plots} on a $201 \times 51$ regular lattice with $\delta = 1$.  Each lattice site is indexed $(i,j)$, for $i = 1,2,\ldots,201$ and $j=1,2\ldots,51$ and is associated with location $(x_i,y_j)$, where $x_i = - 100 + (i-1)\delta$ and $y_j = (j-1)\delta$. Simulations are initialised with one agent on each lattice site in the central region $-10 \le x \le 10$ and all other sites empty. Reflecting boundary conditions are applied at the left and right boundaries of the lattice and periodic boundary conditions are applied at the top and bottom boundaries. The agents initially at $(-10, 25)$, $(0,25)$ and $(10,25)$ are tagged in blue, cyan and green, respectively.  All remaining agents are coloured red.  A single realisation of the non-interacting unbiased random walk model with $P_m=1$ leads to a distribution that appears to be approximately symmetric about $x=0$, and the three tagged agents follow random trajectories with no apparent preferred direction (Figure~\ref{fig:scatter_plots}b).


\begin{figure}
\centering
	\includegraphics[width=0.85\textwidth]{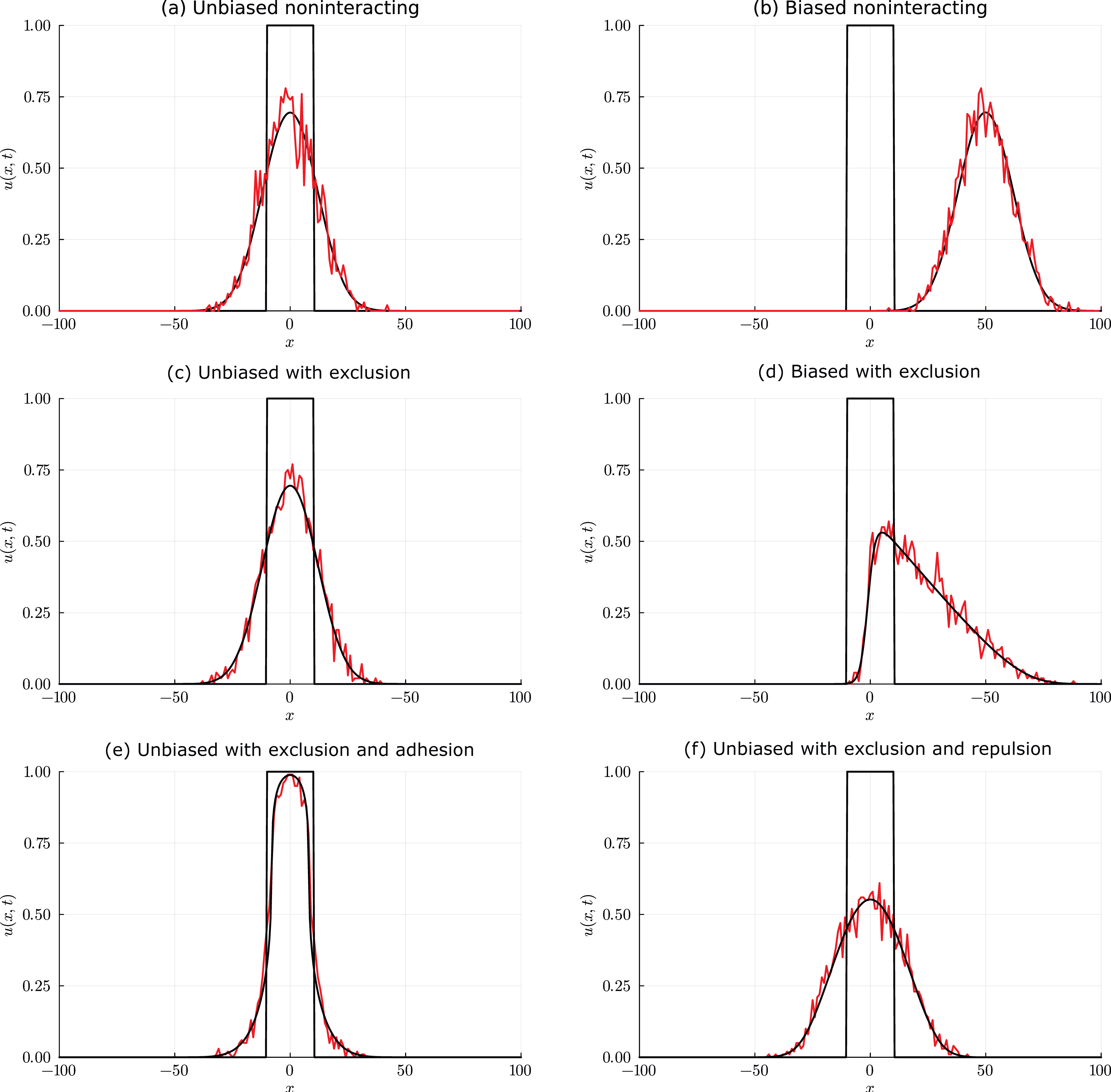}
	\caption{Comparison of column-averaged density data from a stochastic ABM (red) and numerical solution of the corresponding continuum-limit equation (black) for a range of non-interacting and interacting random walk models. All stochastic simulations are performed on a regular $201 \times 51$ lattice with $\delta = 1$, initialised with one agent at each lattice site with $-10 \le x \le 10$ and all other lattice sites empty. All model comparisons are made after 200 time steps of the ABM with $P_m=1$. Results in (a,b) are for a non-interacting random walk. Results in (c,d) are for an interacting exclusion process (i.e.~at most one agent per lattice site). Results in (a,c) are for an unbiased migration mechanism with $P_m=1$ and results in (b,d) are for a biased migration mechanism with $P_m=1$ and $\rho_x =1/2$. The continuum-limit equation for: (a) is the linear diffusion equation, Equation~\eqref{eq:heat}, with $D=1/4$; (b) is the linear advection-diffusion equation, Equation~\eqref{eq:PDE_noninteracting_horizontal_bias}, with $D = 1/4$ and $v_x = 1/4$; (c) is the linear diffusion equation with $D=1/4$; and (d) is the nonlinear advection-diffusion equation, Equation~\eqref{eq:PDE_exclusion_horizontal_bias}, with $D = 1/4$ and $v_x = 1/4$.  Results in (e,f) are for an interacting exclusion process with {\highlight an unbiased migration mechanism} and an adhesion/repulsion mechanism with parameter: (e) $\sigma=3/4$ (adhesion); (f) $\sigma=-3/4$ (repulsion). The continuum-limit equation for (e,f) is Equation~\eqref{eq:PDE_adhesion} with $D=1/4$.}
	\label{fig:pde_plots}
\end{figure}


To quantify the outcome of the stochastic simulations in Figure~\ref{fig:scatter_plots}b, we calculate the average density of agents in each lattice column, $\sum_{j=1}^{51}U_{i,j}/51$, noting that, for the chosen initial condition, the density is independent of $y$. Plotting this quantity as a function of horizontal location $x$ confirms that the density profile is approximately symmetric about $x=0$ (Figure~\ref{fig:pde_plots}a, red). We compare this to solution of Equation~\eqref{eq:heat}, the linear diffusion equation, which we solve in a one-dimensional Cartesian coordinate system since the expected density $p(x,y,t)$ is independent of $y$. The initial condition is $p(x,0)=1$ for $|x| \le 10$ and $p(x,0)=0$ for $|x| > 10$ to match the initial distribution of the column-averaged density of agents, and no-flux boundary conditions are applied at $x=\pm 100$. With $D=P_m\delta^2/(4\tau)=1/4$, we solve Equation~\eqref{eq:heat} numerically using the method of lines~\cite{Morton2005}. In brief, this involves using a standard finite difference approximation for the spatial derivative to discretise the partial differential equation (PDE) on the interval  $-100 \le x \le 100$ using 401 equally-spaced mesh points. This leads to a system of coupled ODEs that we solve numerically using the DifferentialEquations.jl package in Julia. 

Superimposing the numerical solution of Equation~\eqref{eq:heat} (Figure~\ref{fig:pde_plots}a, black) onto the column-averaged density from the ABM reveals a good match between the two density profiles.  The key difference is that the column-averaged density profile involves clearly visible fluctuations whereas the solution of the continuum-limit equation does not. The magnitude of the fluctuations decreases as the vertical height of the lattice increases, or if an ensemble average is taken over multiple independent realisations of the stochastic process. {\highlight Results in Figure~\ref{fig:scatter_plots}c and Figure~\ref{fig:pde_plots}b will be described in Section \ref{sec:biased_random_walks} when we discuss biased lattice-based random walks.}


\subsubsection{Biased lattice-based random walks} \label{sec:biased_random_walks}

In a {\em biased random walk}, the probabilities of moving to each of the adjacent lattice sites are not equal.   This leads to an additional term, called an {\em advection} term or {\em drift} term, in the continuum-limit equation for the mean agent density $p(\vect{x},t)$:
\begin{equation} 
\label{eq:PDE_noninteracting}
\frac{\partial p}{\partial t} = \underbrace{\vphantom{\sum}D\nabla^2 p}_\textrm{random motility} - \underbrace{\vphantom{\sum}\vect{v}\cdot \nabla p,}_\textrm{directed motility}
\end{equation}
where $\vect{v}\in\mathbb{R}^d$ is called the {\em advection velocity} or {\em drift velocity}. Even though movement is now biased in a particular direction, there is still an element of randomness in individual agent movements. This is reflected in the two terms in Equation~\eqref{eq:PDE_noninteracting}: a diffusion term for the random component of motility and an advection term for the directed component. Note that if the advection velocity, $\vect{v}$, is zero, the advection-diffusion equation, Equation~\eqref{eq:PDE_noninteracting}, reduces to the diffusion equation, Equation~\eqref{eq:heat}. Box~1 demonstrates how Equation~\eqref{eq:PDE_noninteracting} along with expressions for $D$ and $\vect{v}$ can be derived from a two-dimensional biased random walk, in the special case where the bias acts only in the horizontal direction so that $\vect{v}=(v_x,v_y)$ with $v_y=0$. {\highlight Note that, although the bias could be thought of as an interaction between the agents and a specified potential field, this is distinct from the agent-agent interactions we will consider in later sections of this article because the agents are still moving independently of each other.}


\bigskip
\begin{tcolorbox}[breakable, title=Box 1: Lattice-based biased random walk model and continuum-limit PDE]
\parskip=6pt

Suppose we have $N$ agents on a two-dimensional square lattice with lattice spacing $\delta$, so that lattice sites have locations $(x,y) = (i \delta, j \delta)$, where $i,j \in \mathbb{Z}$ are standard lattice indices.  In any single realisation of the stochastic ABM, we define $U_{i,j}$ as the number of agents at site $(i,j)$, which is a non-negative, integer-valued random variable. 

Suppose that, during each time step of duration $\tau$, each agent moves with probability $P_m \in [0,1]$, independent of other agents. When an agent at site $(i,j)$ moves, suppose that sites $(i,j \pm 1)$ are chosen with equal probability $1/4$ and sites $(i \pm1 ,j)$ are chosen with potentially unequal probability $(1 \pm \rho_x)/4$, where $\rho_x \in[-1,1]$ is a parameter controlling the bias in the $x$ direction.  

To think about the continuum limit, we use $u_{ij}$ to denote the expected number of agents $\langle U_{i,j}\rangle(t)$ at site $(i,j)$ at time $t$. We can think of this as the average across an {\em ensemble} of $M$ independent, identically-prepared simulations in the limit where $M$ is large:
\begin{equation}
u_{ij} =  \lim_{M\to\infty} \dfrac{1}{M} \sum_{l = 1}^{M}U_{i,j}^{(l)}(t),
\end{equation}
where $U_{i,j}^{(l)}(t)$ is the number of agents at lattice site $(i,j)$ at time $t$ in the $l^\text{th}$ realisation. 

Under these conditions we can write down a conservation statement describing the change in average number of agents as site $(i,j)$ during the time interval from time $t$ to time $t + \tau$, namely
\begin{equation}
\label{eq:conservation_eqn_noninteracting} 
\Delta u_{i,j}  = \dfrac{P_m}{4}\underbrace{\left[\vphantom{\sum}(1+\rho_x)u_{i-1,j}+(1-\rho_x) u_{i+1,j}+ u_{i,j-1}+u_{i,j+1}\right] }_{\textrm{migration onto site $(i,j)$}} - \underbrace{\vphantom{\left[\sum\right]}P_m u_{i,j}.}_{\textrm{migration out of site $(i,j)$}}
\end{equation}
To proceed to the continuum limit, we identify $u_{i,j}(t)/\delta^2$ (i.e. the average density of agents per unit area) with a smooth function $p(x,y,t)$ and expand all terms in Equation~\eqref{eq:conservation_eqn_noninteracting} in a Taylor series about $(x,y) = (i\delta, j\delta)$. Dividing the resulting expressions by $\tau$ and taking the limit as $\delta \to 0$ and $\tau \to 0$ with the ratio $\delta^2/\tau$ held constant, terms of $\mathcal{O}(\delta^3)$ and smaller can be neglected and we are left with
\begin{equation}
\label{eq:PDE_noninteracting_horizontal_bias}
\dfrac{\partial p}{\partial t} =  \underbrace{D\left[\dfrac{\partial^2 p}{\partial x^2} + \dfrac{\partial^2 p}{\partial y^2} \right]}_{\textrm{random motility}} - \underbrace{\vphantom{\left[\dfrac{\partial^2 p}{\partial x^2}\right]}v_x \dfrac{\partial p}{\partial x},}_{\textrm{directed motility}} 
\end{equation}
where
\begin{equation} \label{eq:diffusivity_drift}
D  = \lim_{\substack{\delta \to 0 \\ \tau \to 0}} \left( \dfrac{P_m \delta^2}{4 \tau} \right), \quad  v_x  = \lim_{\substack{\delta \to 0 \\ \tau \to 0}} \left( \dfrac{P_m \rho_x \delta}{2\tau} \right).
\end{equation}
The expressions in Equation~\eqref{eq:diffusivity_drift} relate the parameters in the ABM, $\delta, \tau$, $P_m$ and $\rho_x$, to parameters in the continuum model, $D$ and $v_x$. Note that to obtain a well-defined continuum limit where both $D$ and $v_x$ are $\mathcal{O}(1)$, we have the additional constraint that $\rho_x = \mathcal{O}(\delta)$. In other words, the difference in the probabilities of moving right and left must reduce in proportion to the lattice spacing $\delta$. More generally, there can be bias in the vertical as well as the horizontal direction, leading to  Equation~\eqref{eq:PDE_noninteracting}. 

\end{tcolorbox}
\bigskip


Equations~\eqref{eq:heat} and~\eqref{eq:PDE_noninteracting} are examples of the more general {\em transport equation}
\begin{equation} 
\label{eq:PDE_transport}
\frac{\partial p}{\partial t} = -\nabla \cdot \vect{J},
\end{equation}
where $\vect{J}(\vect{x},t)\in\mathbb{R}^d$ is the {\em flux} of $p$ at $\vect{x}\in R\subseteq\mathbb{R}^d$. In Equation~\eqref{eq:PDE_noninteracting}, the flux is given by $\vect{J}(\vect{x},t) =-D\nabla p+\vect{v}p$, which is the sum of the Fickian diffusive flux ($-D \nabla p$) arising from random motility and the advective flux ($\vect{v}p$) arising from the directional bias. Note that the diffusive flux is directed down the gradient of agent density $\nabla p$, while the advective flux is in the direction of the fixed advection velocity $\vect{v}$. A related type of flux term is a {\em chemotactic flux} $p\chi(c)\nabla{c}$, where $c(\vect{x},t)$ is the chemoattractant concentration and $\chi(c)$ is the chemotactic sensitivity~\cite{othmer1997aggregation}. This type of flux term arises from biased migration up or down the gradient of some chemical species of concentration $c(\vect{x},t)$, the evolution of which may be governed by its own PDE (see~\cite{murray2003mathematical} for more details).

More generally, the probability of movement, $P_m$, may be different for each coordinate direction, such that the probability of moving in the $k^\text{th}$ coordinate direction is $P_{m,k}$. This leads to a diagonal $d\times d$ diffusivity matrix $D$ where $D_{kk}=\lim_{\delta,\tau\to 0} P_{m,k}\delta^2/(2d\tau)$, and hence {\em anisotropic diffusion}. Finally, all the random walk movement parameters can depend on location $\vect{x}$ (in this case the jumps $\vect{Z}_n$ in Equation~\eqref{eq:rw_defn} are no longer IID but depend on $\vect{X}_{n-1}$). In this more general case, a similar procedure to that outlined in Box~1 leads to the following PDE for $p(\vect{x},t)$~\cite{okubo2001diffusion,codling2008random}  
\begin{equation} 
\label{eq:fokker_planck}
\frac{\partial p}{\partial t} =  \underbrace{\vphantom{\sum}\nabla \cdot( D(\vect{x})\nabla p))}_\textrm{random motility} - \underbrace{\vphantom{\sum}\nabla \cdot( \vect{v}(\vect{x}) p)}_\textrm{directed motility},
\end{equation}
where $D(\vect{x})$ is the diffusivity matrix and $\vect{v}(\vect{x})\in\mathbb{R}^d$ is the advection velocity at location $\vect{x}$. Note that Equation~\eqref{eq:fokker_planck} is equivalent to Equation~\eqref{eq:PDE_transport} with flux $\vect{J}(\vect{x})=-D(\vect{x}) \nabla p + \vect{v}(\vect{x}) p$.

Illustrative stochastic simulation results in Figure~\ref{fig:scatter_plots}c show a single realisation of a biased non-interacting random walk on the same regular lattice with the same initial condition used to explore the unbiased non-interacting random walk. This simulation, with $P_m=1$ and $\rho_x = 1/2$, shows that the population of agents moves in the positive $x$-direction, and that the distribution of agents appears to be symmetric about the mean $x$ coordinate of the population, as expected. The trajectories of the three tagged agents show a clear bias in the positive $x$-direction, which is very different to the trajectories for the corresponding unbiased simulation (Figure~\ref{fig:scatter_plots}b). 

Column-averaged density data in Figure~\ref{fig:pde_plots}b confirms that the population density moves in the positive $x$-direction and that the distribution of agent density is approximately symmetric about $x=50$. To compare this to the continuum limit, we solve the linear advection-diffusion equation, Equation~\eqref{eq:PDE_noninteracting_horizontal_bias}, numerically in a one-dimensional Cartesian coordinate system, with $D=P_m\delta^2/(4\tau)=1/4$ and $v_x = P_m\rho_x \delta /(2\tau) = 1/4$, and the same initial and boundary conditions as for the unbiased case. Superimposing the numerical solution onto the column-averaged density data from the ABM reveals a good match between the two density profiles.  We note that the main difference between the results in Figure~\ref{fig:pde_plots}a and those in~\ref{fig:pde_plots}b is the impact of the advection term, which leads to the translation of the population so that the centre of mass of the population (i.e.~the mean agent location) moves at velocity $v_x$.  {\highlight Results in Figure~\ref{fig:scatter_plots}d--f and Figure\ref{fig:pde_plots}c--f will be described later in Sections \ref{sec:crowding} and \ref{sec:other_interactions} where we discuss interacting models.}


\subsection{Lattice-free random walks} 
\label{sec:non_interacting_lattice_free}

ABMs can be implemented as either lattice-based or lattice-free random walks, both of which are encompassed by the definition in Equation~\eqref{eq:rw_defn}. The lattice-based approach has the advantage that simulating movement on a lattice is computationally straightforward and enables easy derivation of the continuum-limit equation (see e.g.~Box~1). However, in biological applications, individuals move in continuous space and are not restricted to an artificial lattice with movement parallel to arbitrary coordinate axes. 

In lattice-free models, agent locations, $\vect{X}_n$, may be any point in $R\subseteq\mathbb{R}^d$ and their direction of movement can be any angle, corresponding to a point on the unit sphere $S^{d-1}$. In the two-dimensional case (i.e.~$d=2$), the direction of movement is specified by an angle $\theta\in [0,2\pi)$, which may be drawn from a suitable circular distribution~\cite{mardia2009directional}. 

If the direction of movement at time $t$ is independent of previous movements, the random walk is referred to as {\em uncorrelated} (see also Section~\ref{sec:position_vs_velocity_jump}). As with lattice-based models, it is useful to derive a continuum-limit equation for the expected agent density as a function of space and time. The continuum-limit equation for an unbiased, uncorrelated lattice-free random walk where the mean squared displacement per step, denoted $\sigma^2 = \langle|\vect{Z}|^2\rangle$, is finite is the linear diffusion equation with diffusivity $D=\lim_{\sigma,\tau\to 0} \sigma^2/(2d\tau)$. This is the same as the lattice-based case (see Section~\ref{sec:fundamentals}). However, for a biased uncorrelated lattice-free random walk, the continuum-limit equation in general depends on how the bias is modelled at the individual level (i.e.~the shape of the distribution of $\vect{Z}_n\in\mathbb{R}^d$) and may include anisotropic diffusion terms~\cite{cheung2007animal,codling2010diffusion}. 


\subsection{Individual-level and population-level behaviour} 
\label{sec:individual_vs_pop_level}

Analytical solutions of the linear advection-diffusion equation, Equation~\eqref{eq:diffusivity_drift}, are available for certain initial and boundary conditions (see e.g.~\cite{montroll1984wonderful,okubo2001diffusion,grimmett2001probability,codling2008random}). Where these are not available, the relevant continuum-limit PDE is usually readily solvable numerically. We do not focus on analytical solutions of Equation~\eqref{eq:diffusivity_drift} in great depth here, as these are typically not applicable in more complex models involving interactions, but we briefly cover some of the general insights that analytical PDE solutions can provide. 

In general, the action of diffusive terms in PDEs is to even out agent density via a net movement of agents from regions of high to low density and corresponding dissipation of density gradients over time (see Figures~\ref{fig:scatter_plots}a-b and~\ref{fig:pde_plots}a). This is analogous to the way heat transfer from hot to cold regions tends to dissipate temperature gradients. The higher the diffusivity, $D$, the faster this process takes place.  At the individual-level, a higher $D$ corresponds to a higher probability of movement, $P_m$, larger average step length, $\delta$, or shorter time step, $\tau$. 

Individual-level measures of movement include the mean location $\langle \vect{X}(t) \rangle$ and mean squared displacement $\langle|\vect{X}(t)-\vect{X}(0)|^2\rangle$ at time $t$. For an unbiased random walk, the mean location is always equal to the starting location $\vect{X}(0)=\vect{x}_0$ and, provided the mean squared displacement per step $\sigma^2$ is finite, the mean squared displacement at time $t$ is $\langle|\vect{X}(t)-\vect{X}(0)|^2\rangle=2dDt$~\cite{berg1983random}. Equivalently, $\langle|\vect{X}_n-\vect{X}_0|^2\rangle=n \sigma^2$ in terms of the individual-level variables.  

The fact that the mean of the {\em squared} distance from the starting location is linear in $t$ is a fundamental property of random undirected motility and the diffusive process to which it converges in the continuum limit. Intuitively, this occurs because agents undergoing an unbiased random walk do not take a series of steps in the same direction (which would lead to the mean squared displacement increasing with $t^2$) but typically take a tortuous path with frequent backtracking. For random walks where $\langle |\vect{Z}_n|^2 \rangle$ is infinite, the step length distribution is referred to as {\em heavy-tailed}, and the mean squared displacement scales asymptotically as $t^\nu$ with $1<\nu<2$, which is a type of anomalous diffusion~\cite{weeks1996anomalous}. This is true of power-law distributions for $\vect{Z}_n$ with power $\alpha<3$, which lead to L\'evy walks (see e.g.~\cite{edwards2007revisiting}). We do not consider these further here and refer the reader to~\cite{viswanathan2011physics} for more details.   

Advection terms act to translate the density profile $p(\vect{x},t)$ through space, with a direction and speed determined by the advection velocity $\vect{v}=(v_x,v_y)$, without changing its shape (compare Figures~\ref{fig:pde_plots}a and b). At the individual-level, a stronger directional bias (i.e.~a bigger difference between the probabilities of moving in opposing directions) leads to a faster advection speed. For the biased random walk described by Equation~\eqref{eq:PDE_noninteracting}, the mean location is $\langle \vect{X}(t)\rangle=\vect{x}_0+\vect{v}t$ and the mean squared displacement is $\langle |\vect{X}(t)-\vect{X}(0)|^2\rangle=|\vect{v}|^2t^2 + 2dDt$~\cite{codling2008random}. The quadratic term in $t$ here indicates that the population is moving ballistically rather than purely diffusively, reflecting the effect of the bias. However, the variance in agent location, which measures dispersal about the mean location, increases linearly with $t$, as for the unbiased case~\cite{codling2008random}. Hence, a biased lattice-based random walk can be viewed as an isotropic diffusive process taking place in a moving reference frame.

Both diffusion and advection terms preserve total population size $\int_R p(\vect{x},t)\,\mathrm{d}\vect{x}$. If there is proliferation or death, this introduces a {\em reaction term} into the continuum-limit PDE, which causes the total population size to change over time (see Section~\ref{sec:birth_death}). 


\subsection{Position-jump and velocity-jump processes} 
\label{sec:position_vs_velocity_jump}

Sections~\ref{sec:fundamentals}--\ref{sec:individual_vs_pop_level} concern random walks that are Markov processes with respect to the agent's location $\vect{X}_n$. These are known as {\em position-jump processes} because agent positions change over time via a series of discrete jumps. Another type of random walk is a {\em velocity-jump process}, which is a Markov process with respect to the agent's velocity $\vect{V}$. The agent's location changes according to $\mathrm{d}\vect{X}/\mathrm{d}t=\vect{V}$, while the velocity changes via a series of discrete jumps, which can occur at fixed or variable time steps~\cite{kareiva1983analyzing,othmer1988models,othmer2000diffusion}. {\highlight Velocity-jump processes are closely related to a class of models called active-particle models in the physics literature, which consist of systems of coupled nonlinear Langevin equations \cite{bellomo2017active}. This subsection gives a very brief outline of velocity-jump processes, but for the remainder of the article we focus mainly on position-jump processes. }

Velocity-jump processes generally describe {\em correlated random walks}, meaning that the velocity after $n$ jumps is correlated with the velocity after $n-1$ jumps. They are useful to model agents that have a tendency to continue moving in the same or similar direction, a feature known as persistence~\cite{patlak1953random}. Velocity-jump processes have applications in a host of areas, including swimming micro-organisms~\cite{hill1997biased}, bacterial chemotaxis~\cite{berg1983random,bearon2000modelling,othmer2002diffusion}, angiogenesis~\cite{stokes1991analysis,plank2004lattice}, animal movement~\cite{kareiva1983analyzing,bovet1988spatial,codling2004random} and navigation~\cite{cheung2007animal,codling2007group}.

The natural setting for velocity-jump processes where agent velocities with arbitrary directions are allowed is a lattice-free model (see Section~\ref{sec:non_interacting_lattice_free}). The general continuum-limit for velocity-jump processes is the {\em linear transport equation}, which is a PDE for the density $p(\vect{x},\vect{v},t)$ of agents that have location $\vect{x}$ and velocity $\vect{v}$ at time $t$~\cite{othmer1988models}. {\highlight When applied to particles undergoing Brownian motion, this gives rise to the {\em Klein-Kramers} equation \cite{fa2013generalized}. } The simplest example of a velocity-jump process is the one-dimensional case where agents have fixed speed $s>0$ and can either be moving right ($v=s$) or left ($v=-s$). In this case, it is possible to derive a continuum-limit for $p(x,t)$ called the {\em telegraph equation}~\cite{goldstein1951diffusion,kac1974stochastic}. The telegraph equation can be generalised to higher dimensions, however this requires that agent velocities at any given time are aligned with one of the coordinate axes.

One limitation of the advection-diffusion equations introduced in Section~\ref{sec:non_interacting_lattice_based} that arise from position-jump processes is that they have an infinite propagation speed~\cite{okubo2001diffusion}. For example, the solution $p(\vect{x},t)$ to Equation~\eqref{eq:PDE_noninteracting} is strictly positive for all locations $\vect{x}$ at any $t>0$~\cite{codling2008random}. This implies that an agent has non-zero probability of being arbitrarily far away from its starting location at arbitrarily small times. This is a consequence of the fact that the continuum limit requires holding $\delta^2/\tau$ constant as $\delta,\tau\to 0$ which, in turn, implies that the movement speed is unbounded because $\delta/\tau\to\infty$ as $\delta,\tau\to 0$. In reality, agents can only move at finite speeds and so advection-diffusion PDEs such as Equation~\eqref{eq:PDE_noninteracting} should be viewed as an approximation that is valid after a sufficiently large number of random walk steps. Velocity-jump processes overcome this limitation by assigning agents a finite velocity. Transport equations derived from velocity-jump processes typically tend asymptotically to advection-diffusion equations as $t\to\infty$, because the short-term correlations in agent velocities become negligible at large times~\cite{othmer1988models,othmer2000diffusion}. For example, reaction-diffusion equations for chemotactic movement can be derived as the diffusive, long-time limit of a velocity-jump process~\cite{erban2004individual,erban2005signal}. {\highlight Similarly, velocity-jump processes can be asymptotically analysed under different scalings to arrive at different forms of the PDE model~\cite{othmer1988models,othmer2000diffusion,fa2013generalized}.} 


\section{Including interactions: crowding effects} 
\label{sec:crowding}

All the models described in Section~\ref{sec:non_interacting} assume that agents behave completely independently of one another. In reality, the behaviour of individual agents may be influenced by other agents in various ways. One important example is that no two agents can occupy the same physical space, which we refer to as {\em volume exclusion}. A related example is that agents may be in competition for resources, meaning that the presence of nearby agents may reduce an agent's proliferation rate or increase its death rate. Collectively, we refer to these types of interactions as {\em crowding effects}. There are various ways of incorporating crowding effects into models, and the choice of model will be influenced by the underlying biology. For example, can agents be approximated as hard, rigid objects of a fixed size, or can they deform or be compressed by the presence of neighbouring agents? Is the direction of movement always random or do agents preferentially move towards or away from neighbouring agents?

In Section~\ref{sec:birth_death}, we consider models that include agent proliferation and death. Including density-dependence in proliferation or death rates is one way to account for local competition, but does not enforce volume exclusion if motility is still independent of other agents. In Section~\ref{sec:crowding_lattice_based}, we show how volume exclusion can be incorporated into lattice-based models by disallowing movement to a lattice site that is already occupied. In Section~\ref{sec:crowding_lattice_based_with_proliferation}, we generalise this to include populations with proliferation. Sections \ref{sec:crowding_lattice_free}--\ref{sec:crowding_lattice_free_with_proliferation} extends these ideas to lattice-free models. As in Section~\ref{sec:non_interacting}, we focus on the derivation of continuum-limit approximations of discrete ABMs where possible. We will highlight that care needs to be taken to ensure the interactions between agents are correctly captured within the limiting process. We will also point out situations where the continuum-limit approximation breaks down and an alternative approach is needed. {\highlight There are alternative approaches to obtain continuous models for particle density departing from discrete individual-based descriptions, for example methods based on the Doi-Peliti formalism \cite{doi1976stochastic,peliti1985path,isaacson2008relationship}, but we do not consider these further here. }


\subsection{Proliferation and death} 
\label{sec:birth_death}

In the models considered in Section~\ref{sec:non_interacting}, the population always had a fixed number of agents, $N$. In many applications, agents can proliferate and/or die during the time period of interest. This can be incorporated into ABMs by assigning each agent a probability of proliferating or dying in each time step, along with rules specifying how the locations of {\em daughter agents} are determined relative to the parent. In general, this leads to an additional term in the continuum-limit PDE for mean agent density $p(\vect{x},t)$. For example, when the motility terms are written in the form of a general transport equation with flux $\mathbf{J}$, as in Equation~\eqref{eq:PDE_transport}, the PDE becomes
\begin{equation} 
\label{eq:PDE_transport_reaction}
\frac{\partial p}{\partial t} = -\nabla \cdot \vect{J}  + F(p).
\end{equation}
The term $F(p)$ is called a {\em reaction term}, also known as a {\em source term}, and represents the local net rate of change in expected agent density due to proliferation and death. 

The expected total population size $N(t)=\int_R p(\vect{x},t) \, \textrm{d}\vect{x}$ is now not fixed but time-dependent. A differential equation for $N(t)$ may be obtained by integrating Equation~\eqref{eq:PDE_transport_reaction} over the domain $R$ and applying the divergence theorem:
\begin{equation} \label{eq:pop_size_conservation}
    \frac{\text{d}N}{\text{d}t} = -\int_{\partial R} \vect{J} \cdot \mathrm{d}\hat{\vect{n}} + \int_R F(p(\vect{x},t)) \, \mathrm{d}\vect{x}. 
\end{equation}
{\highlight The first integral on the right-hand side of Equation~\eqref{eq:pop_size_conservation} represents the net flux $\vect{J}$ of agents out of the domain (i.e. total outward flux minus total inward flux)} across the boundary $\partial R$, whose unit outward normal vector is denoted $\hat{\vect{n}}$. This will be determined by the boundary conditions on Equation~\eqref{eq:PDE_transport_reaction} and it will be zero in the case of periodic or no-flux boundary conditions. The second term in Equation~\eqref{eq:pop_size_conservation} represents the net aggregate population growth rate due to proliferation and death. 

The simplest example is where each agent has fixed probabilities of proliferation $P_p$ and death $P_d$ in each time step $\tau$, and daughter agents are initially placed on the same lattice site as their parent. This is referred to as {\em density-independent} proliferation and death, and this model does not include any interactions among agents. The reaction term is $F(p)=(b-\mu)p$, where $b=\lim_{\tau\to 0} P_p/\tau$ and $\mu=\lim_{\tau\to 0} P_d/\tau$ are, respectively, the proliferation rate and death rate per unit time. In this case, Equation~\eqref{eq:PDE_transport_reaction} with a diffusive flux $\vect{J}=-D\nabla p$ is known as {\em Skellam's equation}~\cite{skellam1951dispersal}. The net difference $r=b-\mu$ between the proliferation rate and the death rate is referred to as the {\em intrinsic population growth rate}.  It follows from Equation~\eqref{eq:pop_size_conservation} that, if there is no net flux of agents {\highlight across the domain boundary}, the total population size $N(t)$ grows exponentially if $r>0$ and decays exponentially if $r<0$. Note that for the proliferation and death rates to be finite in the continuum limit requires that the probabilities $P_p$ and $P_d$ in the ABM are $\mathcal{O}(\tau)$ as $\tau\to 0$.

One way to generalise this simple example to include crowding effects is to allow the probability of death, $P_d$, to depend on the number of agents $U_{i,j}$ occupying the lattice site. This model does not include volume exclusion as multiple agents per lattice site are still allowed, but does include the effect of local competition. {\highlight Suppose we set $P_d=C_1+C_2 U_{i,j}/\delta^2$, so that the probability of death increases with local agent density $U_{i,j}/\delta^2$. Then the reaction term becomes $F(p)= (b-\mu_1)p - \mu_2 p^2$, where $\mu_1=\lim_{\tau\to 0} C_1/\tau$ and $\mu_2=\lim_{\tau\to 0} C_2/\tau$ are the rates of density-independent and density-dependent death, respectively.} This is known as the {\em logistic growth function} and, in this case, Equation~\eqref{eq:PDE_transport_reaction} with a diffusive flux $\vect{J}=-D\nabla p$ is known as the {Fisher-KPP equation}, more commonly written as 
\begin{equation} \label{eq:fisher}
    \frac{\partial p}{\partial t} = \underbrace{D\nabla^2 p \vphantom{\left( \frac{p}{K}\right)} }_\textrm{random motility}  + \underbrace{rp\left(1-\frac{p}{K}\right)}_\textrm{net proliferation},
\end{equation}
where $r=b-\mu_1$ is the intrinsic population growth rate and $K=(b-\mu_1)/\mu_2$ is the {\em carrying capacity density}. Note that the density-dependence in the death rate leads to a nonlinear term in the continuum limit, Equation~\eqref{eq:fisher}. The combination of random undirected motility and density-dependent growth in Equation~\eqref{eq:fisher} leads to {\em travelling wave} behaviour for certain initial conditions~\cite{murray2003mathematical}. The diffusion term causes the population spread out in space, allowing the wavefront to advance into previously unoccupied areas, while the reaction term causes the density behind the wavefront to grow asymptotically towards the carrying capacity density $K$.  

The Fisher-KPP equation and related nonlinear reaction-diffusion equations can used to model motile populations with density-dependent proliferation and death. However, they do not take account of the effects of agent interactions on motility, which is still assumed to be independent of other agents. In Section~\ref{sec:crowding_lattice_based}, we show how the effects of crowding on motility can be incorporated via volume exclusion.


\subsection{Lattice-based models with volume exclusion} 
\label{sec:crowding_lattice_based}

One of the simplest ways to incorporate crowding effects into lattice-based models is to assume that each lattice site is the size of a single agent and that it can contain at most one agent. An ABM where agents undergo random walks with volume exclusion is referred to as an {\em exclusion process}, {\highlight which is closely related to a class of models known as contact processes and voter process (see \cite{durrett1994stochastic,liggett1999stochastic}), although we do not discuss these further here}. In an exclusion process, if an agent attempts to move onto a lattice site that is already occupied, the move is aborted. In deriving a corresponding continuum-limit PDE, we need to take into account the fact that the probability of an agent moving from its current site to an adjacent site depends on whether or not the target site is occupied. 

In Box~2, we show how an approximate continuum-limit equation for agent density can be derived for a two-dimensional, lattice-based exclusion process with bias in the horizontal direction  \cite{Simpson2009}. This takes the form of a nonlinear advection-diffusion PDE.
In the more general $d$-dimensional case with bias vector $\vect{v}\in\mathbb{R}^d$, this PDE may be written
\begin{equation} \label{eq:PDE_exclusion}
    \frac{\partial u}{\partial t} = \underbrace{\vphantom{\sum}D\nabla^2 u}_\textrm{random motility} - \underbrace{\vphantom{\sum}\vect{v}\cdot \nabla\left( u(1-u)\right)}_\textrm{directed motility},
\end{equation}
{\highlight where $u$ is the {\em dimensionless agent density}, which is equivalent to the density relative to that of a fully occupied lattice where $u=1$.} Unlike in the non-interacting case in Section~\ref{sec:non_interacting_lattice_based}, when a lattice is used to model volume exclusion, the lattice spacing $\delta$ must correspond approximately to the size of an agent. It may seem contradictory then that Equation~\eqref{eq:PDE_exclusion} is used to model agents of a fixed size $\delta>0$, yet is derived by taking the limit $\delta\to 0$. However, as for the non-interacting case, Equation~\eqref{eq:PDE_exclusion} should be interpreted as an {\em approximation} to the discrete ABM with fixed $\delta>0$, which is valid after a sufficiently large number of steps. The size of agents, $\delta$, enters into Equation~\eqref{eq:PDE_exclusion} indirectly because the agent density per unit volume, $p$, is related to $u$ via $p=u/\delta^d$, and so under volume exclusion the maximum value of $p$ is $1/\delta^d$.



\bigskip
\begin{tcolorbox}[breakable, title=Box 2: Lattice-based biased random walk and continuum-limit PDE with exclusion]
\parskip=6pt

Suppose we have the same lattice as in Box~1, except now in any realisation of the ABM the occupancy status $U_{i,j}$ of lattice site $(i,j)$ is a binary random variable with $U_{i,j} = 1$ if the site is occupied (by at most a single agent) and $U_{i,j}=0$ if the site is vacant. Unlike in the non-interacting random walk, the order in which agents attempt to move matters because when one agent moves, it changes which sites are available/occupied for other agents. To deal with this we use a {\em random sequential update} method: during each time step $\tau$, $N$ agents are selected uniformly at random, one at a time with replacement. When an agent at site $(i,j)$ is selected, it attempts to move with probability $P_m$, and chooses target sites $(i,j \pm 1)$ with equal probability $1/4$ and sites $(i\pm1,j)$ with potentially unequal probability $(1\pm\rho_x)/4$ where $\rho_x\in[-1, 1]$, as in Box~1. If an agent attempts to move to a target site that is already occupied, the move is aborted. 

Considering an ensemble of $M$ identically prepared realisations, we can write down an approximate conservation statement describing the change in expected occupancy $u_{i,j}=\langle U_{i,j}\rangle$ of site $(i,j)$ during a time step $\tau$, namely
\begin{align}
&\Delta u_{i,j}  = \dfrac{P_m}{4}  \underbrace{\left(1-u_{i,j}\right)\left[(1+\rho_x)u_{i-1,j}+(1-\rho_x) u_{i+1,j}+ u_{i,j-1}+u_{i,j+1}\right] }_{\textrm{migration onto site $(i,j)$}} \label{eq:conservation_eqn_exclusion} \\
&-\dfrac{P_m}{4}\underbrace{u_{i,j}\left[(1+\rho_x)\left(1-u_{i+1,j}\right) +(1-\rho_x)\left(1-u_{i-1,j}\right) +\left(1-u_{i,j-1}\right) +\left(1-u_{i,j+1}\right)  \right]}_{\textrm{migration out of site $(i,j)$}}, \notag  
\end{align}
where we see the factors $(1-u)$ appearing in the transition probabilities, which approximately ensures that motility events require the target site to be vacant. Whilst the corresponding continuum-limit equation for the non-interacting case, Equation~\eqref{eq:conservation_eqn_noninteracting},  is exact, Equation~\eqref{eq:conservation_eqn_exclusion} is approximate as it assumes the random variables representing the occupancy status of adjacent sites are independent. This allows expressions of the form $\langle U_{i,j}U_{i\pm 1,j} \rangle$ to be approximated as $u_{i,j}u_{i\pm 1,j}$.

We identify $u_{i,j}$ with a smooth function $u(x,y,t)$ and expand all terms in Equation~\eqref{eq:conservation_eqn_exclusion} as Taylor series about $(x,y) = (i\delta, j\delta)$. Dividing the resulting expressions by $\tau$ and taking the limit as $\delta \to 0$ and $\tau \to 0$ with the ratio $\delta^2/\tau$ held constant gives
\begin{equation}
\label{eq:PDE_exclusion_horizontal_bias}
\dfrac{\partial u}{\partial t} =  \underbrace{D\left[\dfrac{\partial^2u}{\partial x^2} + \dfrac{\partial^2u}{\partial y^2} \right]}_{\textrm{random motility}} - \underbrace{\vphantom{\left[\dfrac{\partial^2u}{\partial x^2}\right]}v_x \dfrac{\partial}{\partial x} \left[u(1-u) \right],}_{\textrm{directed motility}} 
\end{equation}
where
\begin{equation}
D  = \lim_{\substack{\delta \to 0 \\ \tau \to 0}} \left( \dfrac{P_m \delta^2}{4 \tau} \right), \quad  v_x  = \lim_{\substack{\delta \to 0 \\ \tau \to 0}} \left( \dfrac{P_m \rho_x \delta}{2\tau} \right).
\end{equation}
Note that there can be bias in the vertical as well as the horizontal direction, leading to the more general nonlinear advection-diffusion equation stated in Equation~\eqref{eq:PDE_exclusion}.

\end{tcolorbox}
\bigskip


The derivation of Equation~\eqref{eq:PDE_exclusion} requires an assumption that the random variables representing the occupancy status of any two adjacent lattice sites are independent \cite{Simpson2009}. This is known as the {\em mean-field assumption} and allows the probability that two neighbouring sites are occupied to be approximated as the product of the two single-site occupancy probabilities. This assumption is reasonable in the absence of mechanisms, such as rapid proliferation or neighbour-dependent death, that generate significant spatial correlations between agent locations. However, in case where the population does have strong spatial structure, the approximation breaks down and an alternative approach is required (see Section~\ref{sec:spatial_structure_lattice_based}). {\highlight The mean-field approximation tends to break down particularly readily in one spatial dimension, and a range of analytical results have been derived for one-dimensional exclusion processes, sometimes referred to as {\em single-file diffusion} models \cite{wei2000single,schonherr2004exclusion,lakatos2006hydrodynamic,grabsch2024tracer}. However, in this review we focus primarily on two- and three-dimensional models as they are more  relevant to a broader class of experimental observations.}

Interestingly, in the absence of bias in agent movements, $\vect{v}=\vect{0}$ and the continuum-limit PDE for the exclusion process in Equation~\eqref{eq:PDE_exclusion} is simply the linear diffusion equation, as is the case for non-interacting agents. This means that at the population-level it is impossible to distinguish whether crowding interactions are taking place between agents or not. On the other-hand, when individual agent movements are biased, Equation~\eqref{eq:PDE_exclusion} differs from the continuum-limit equation for a population of non-interacting agents, Equation~\eqref{eq:PDE_noninteracting}, because the advection term is nonlinear. For a population of non-interacting agents, the advective flux is $\vect{v}u$ whilst for the exclusion process it is $\vect{v}u(1-u)$. At low densities, the predictions of the two different models will be similar as $u(1-u) \approx u$ for $u \ll 1$. This makes sense because it is relatively rare for two agents to occupy adjacent lattice sites when the density is low.

Intuitively, the reason volume exclusion does not affect the continuum limit for random undirected motility is that attempted movements in one direction that are aborted due to the target site being occupied occur with the same frequency as aborted movements in the opposite direction. Thus, although there are fewer movements overall in the exclusion process than in the non-interacting random walk, the net effect of volume exclusion on the diffusive flux at any location $\vect{x}$ is zero. In contrast, when migration is biased, attempted movements in one direction are more likely than in the opposite direction. This means that a reduction in movement due to crowding translates into a reduction in the net advective flux in the continuum-limit equation. Thus, crowding does not affect diffusion, but slows advection in regions of high agent density.

Figure~\ref{fig:scatter_plots}e,f shows example simulations of lattice-based random walks with exclusion, with the same lattice and initial condition as previously. A single realisation of the unbiased exclusion process with $P_m=1$ and $\rho_x=0$  (Figure~\ref{fig:scatter_plots}e) leads to a population-level distribution that is difficult to distinguish from the unbiased non-interacting simulation (Figure~\ref{fig:scatter_plots}b). However, the trajectories of the tagged agents appear very different from the non-interacting case. The agent initially at the left-most leading edge drifts in the negative $x$-direction, the agent initially at the right-most leading edge drifts in the positive $x$-direction, and the agent initially in the centre of the population hardly moves at all during the simulation. This shows that crowding affects the movements of individual agents, by making movement less likely in some or all directions, but that it does not affect the population-level distribution~\cite{SimpsonPathlines}. 

The biased exclusion process simulation with $P_m=1$ and $\rho_x = 0.5$ (Figure~\ref{fig:scatter_plots}f) shares some similarities with the biased non-interacting simulation (Figure~\ref{fig:scatter_plots}c) since the population of agents drifts in the positive $x$-direction. Again, we see differences in the trajectories of tagged agents since the agent initially at the right-most leading edge drifts in the positive $x$-direction, whereas the agents in the centre and at the left-most leading edge barely move at all during the simulation. This is very different to the biased non-interacting model where all three tagged agents exhibit a clear drift in the positive $x$-direction over time.

Column-averaged density data from the exclusion process simulations show a good match with numerical solutions of the continuum-limit PDE, Equation~\eqref{eq:PDE_exclusion_horizontal_bias}. The continuum-limit equation for the unbiased exclusion process model (Figure~\ref{fig:pde_plots}c) is identical to that for the unbiased non-interacting model  (Figure~\ref{fig:pde_plots}a). The continuum-limit equation for the biased exclusion process model  (Figure~\ref{fig:pde_plots}d) involves a nonlinear advection term, leading to a solution that is qualitatively different to the biased non-interacting model  (Figure~\ref{fig:pde_plots}b). The density profile for the biased exclusion process is asymmetric and the aggregate net drift in the positive $x$ direction is significantly less than in the non-interacting case. This is because the crowding effects have little impact on the right-hand leading edge of the population, where there is a relatively low probability of movement being impeded, but have a strong effect both in the centre and at the left-hand leading edge of the population, where there is a high probability of being impeded and consequent reduction in advection. Together, the suite of continuum--discrete comparisons in Figure~\ref{fig:pde_plots}a-d confirms that population-level density information for unbiased motility is insensitive to whether the model includes volume exclusion or not, whereas population-level density information for biased motion is strongly impacted by crowding effects.

One way to distinguish between models with and without volume exclusion is to consider how the mean displacement and mean squared displacement of individual agents on the two-dimensional lattice evolve over time. In the non-interacting case, the mean displacement satisfies $\text{d}\langle \vect{X}(t)\rangle/\text{d}t=\vect{v}$ in the continuum limit, i.e.~agents are simply advected with constant velocity $\vect{v}$ (as seen in Section~\ref{sec:individual_vs_pop_level}). In the exclusion process we have $\text{d}\langle \vect{X}(t)\rangle/\text{d}t=-2D\nabla u +\vect{v}(1-u)$, which shows that agents both move down gradients in local density $u$ and are advected at a rate proportional to $1-u$, which is the fraction of unoccupied space. Similarly, in the unbiased non-interacting case, the mean squared displacement satisfies $\text{d}\langle |\vect{X}(t)|^2 \rangle/\text{d}t = 4D$, whilst in the unbiased exclusion process we have $\text{d}\langle |\vect{X}(t)|^2 \rangle/\text{d}t = 4D\left(1-u-\langle \vect{X}(t)\rangle \cdot \nabla u\right)$. The rate of increase of mean squared displacement with time is sometimes referred to as the {\em self-diffusivity} of an individual agent. As we have seen, the {\em collective diffusivity}, as measured by the diffusion coefficient $D$ in the continuum-limit equation, is the same for the non-interacting random walk and the exclusion process. However, the expressions above for  $\text{d}\langle |\vect{X}(t)|^2 \rangle/\text{d}t$ show that crowding effects reduce the self-diffusivity of an individual agent as the local density $u$ increases, at least for agents that are not on a steep gradient in $u$. These differences can be clearly seen in the tagged agent trajectories depicted in Figure~\ref{fig:scatter_plots}.


\subsection{Lattice-based models with volume exclusion, proliferation and death}  \label{sec:crowding_lattice_based_with_proliferation}

It is simple to generalise the derivation in Box~2 to include agent proliferation. Suppose that, in addition to attempting to move with probability $P_m$ per time step of duration $\tau$, agents die with probability $P_d$ and attempt to proliferate with probability $P_p$, with the daughter agent placed on a randomly selected adjacent lattice site. If the selected site is already occupied, the proliferation attempt is aborted. This results in the continuum-limit PDE
\begin{equation} 
\label{eq:PDE_exclusion_proliferation}
\dfrac{\partial u}{\partial t} = \underbrace{\vphantom{ru(1-u)}D\nabla^2 u}_\textrm{random motility} - \underbrace{\vect{v}\cdot \nabla  \left(u(1-u)\right)}_\textrm{directed motility} + \underbrace{ru\left(1-\frac{u}{K}\right)}_\textrm{proliferation \& death},
\end{equation}
where $r=\lim_{\tau\to0}((P_p-P_d)/\tau)$ and $K=1-P_d/P_p$. Note that this is equivalent to the Fisher-KPP equation, Equation~\eqref{eq:fisher}, with an additional nonlinear advection term. In the case where there is no death (i.e. $P_d=0$), the carrying capacity density $K$ equals $1$, corresponding to a fully occupied lattice. If $P_d>0$ then the carrying capacity density $K$ is less than $1$. 

Comparisons between stochastic simulations of the ABM and the solution of Equation~\eqref{eq:PDE_exclusion_proliferation} confirm that Equation~\eqref{eq:PDE_exclusion_proliferation} provides a good approximation of the dynamics of the ABM provided that proliferation events occur infrequently relative to movement events (i.e.~$P_p/P_m \ll 1$)~\cite{Simpson2010}.  If $P_p/P_m$ is too large, daughter agents tend to remain close to their ancestors for some time in the ABM, which leads to the formation of clusters of agents~\cite{baker2010correcting}. This violates the mean-field assumption and an alternative approach is required (see Section~\ref{sec:spatial_structure_lattice_based}).


\subsection{Lattice-free models with volume exclusion} 
\label{sec:crowding_lattice_free}

As shown in Section~\ref{sec:crowding_lattice_based}, a lattice-based model that allows at most one agent per lattice site imposes a maximum density on the population that is determined by the lattice spacing. However, this maximum density can only be realised if neighbouring agents are perfectly aligned on a regular lattice. This is not usually biologically realistic as, in reality, there is nothing that confines agents to a lattice and as a result agents tend to be positioned more irregularly. For example, experimental imaging of {\em in vitro} cell populations (see e.g.~\cite{khain2011collective}) reveals that, even as populations grow to a limiting density (i.e.~carrying capacity), there are typically gaps between cells and they are not able to efficiently fill space. Thus, whilst lattice-based models may be a reasonable approximation for populations at relatively low density, for populations with high or growing density they make unrealistic assumptions about how neighbouring agents are aligned. This may result in inaccurate predictions of the movement of high-density populations, and overestimate the maximum density that a proliferative population can grow to. 

Initially, most biological applications of lattice-free models ignored interactions between agents~\cite{othmer2000diffusion,coscoy2007statistical,ziff2009capture} or were primarily simulation-based~\cite{stokes1991analysis,plank2004lattice,codling2007group}. However, increasingly, efforts have been made to derive approximate descriptions of agent density in lattice-free models that include agent-agent interactions such as crowding effects. 

Dyson et al.~\cite{dyson2012macroscopic} derived a nonlinear diffusion equation for the random motility of agents in a one-dimensional lattice-free model. They showed that volume exclusion leads to a nonlinear diffusion equation
\begin{equation}
\frac{\partial u}{\partial t} = \nabla \cdot \left( D(u) \nabla u\right),
\end{equation}
with density-dependent collective diffusivity $D(u)$. This contrasts with the lattice-based case, where random motility leads to linear diffusion (i.e.~constant diffusivity, $D$) as seen in Equation~\eqref{eq:PDE_exclusion_horizontal_bias}. In the lattice-free model, $D(u)$ is an increasing function of density when the step size is small relative to the agent radius, because volume exclusion means more movements occur in the direction of unoccupied space. Conversely, if the step size is sufficiently large relative to agent radius, $D(u)$ decreases with $u$ because a large proportion of attempted moves are aborted~\cite{dyson2012macroscopic}. This approach was later extended to higher dimensions~\cite{dyson2015importance}. In the limit where the step size is small and the population size is large, the collective diffusivity is given by $D(u)=D_0\left( 1+4(d-1)u\right)$ when expressed in terms of the density $u\in[0,1]$, where $D_0$ is the diffusivity at low density and $d\in\left\{2,3\right\}$ is the number of spatial dimensions.   

Bruna and Chapman~\cite{bruna2012excluded} also considered the diffusive movement of agents with volume exclusion. Identifying the agent radius relative to the domain size as a small parameter $\epsilon$, they used matched asymptotic expansions to derive a continuum approximation for the mean density of agents at location $\vect{x}$ and time $t$ to leading order in $\epsilon$. As in~\cite{dyson2015importance}, the resulting PDE is a nonlinear diffusion equation with collective diffusivity $D(u)=D_0\left( 1+4(d-1)u\right)$ in the limit of a large population size. The same methodology was also applied to similar models with multiple species of agents~\cite{bruna2012diffusion} and agents moving in confined geometries~\cite{bruna2014diffusion}. This approach has the advantage of accounting for correlations in agent locations (see also Section~\ref{sec:spatial_structure}). However, it is only valid for sufficiently small $\epsilon$, meaning that the total volume occupied by all agents combined is small relative to the size of the domain. 

In contrast to position-jump processes such as those described above, velocity-jump processes have primarily been used in cases where the agents are assumed not to interact. This means that agents can be arbitrarily close to one another and their movements are independent of the locations of other agents~\cite{patlak1953random,goldstein1951diffusion}.  More recent work has explored the consequences of including volume exclusion in a one-dimensional setting~\cite{Treloar2011}, where it is possible to introduce different types of crowding effects. These different mechanisms have subtly different outcomes in terms of the resulting continuum-limit equation. {\highlight Although some approximate results have been derived in the active matter context (see \cite{alert2020physical} and references therein), deriving mean-field equations for velocity-jump processes with crowding effects in higher dimensions is relatively under-explored in comparison to the position-jump case.}


\subsection{Lattice-free models with volume exclusion and proliferation}  \label{sec:crowding_lattice_free_with_proliferation}

The examples in Section~\ref{sec:crowding_lattice_free} relate to lattice-free models of populations with volume exclusion but without proliferation. In a volume exclusion model where agents can proliferate, derivation of a continuum-limit approximation needs to account for the fact that proliferation requires sufficient space for the daughter agent. This is more complex in lattice-free models than lattice-based models because the space required for proliferation has a non-trivial geometry and depends on the direction of dispersal of the daughter agent. Furthermore, unlike non-proliferative models, proliferation leads to the possibility that agent density can increase to high levels over time, making crowding effects stronger. For lattice-based exclusion models, including proliferation is relatively straightforward as the probability of successful proliferation can be expressed in terms of the probability of occupancy of adjacent lattice sites (see Sections \ref{sec:crowding_lattice_based}--\ref{sec:crowding_lattice_based_with_proliferation}). However, for lattice-free models this simplification is not available and the task of deriving an approximation for the growth function $F(u)$ in terms of local agent density $u$ is more complicated. 

Plank and Simpson~\cite{plank2012models} derived an approximation for the growth function $F(u)$ that gives a better match with simulations of a spatially uniform lattice-free ABM than the standard logistic growth function. As with the continuum-limit equation for a lattice-based model of a proliferative population in Equation~\eqref{eq:PDE_exclusion_proliferation}, this approximation requires the probability of proliferation to be low relative to the probability of movement so that clusters of agents do not develop (see Section~\ref{sec:spatial_structure_lattice_free}). Plank and Simpson showed that the modified growth equation led to a different interpretation of experimental data and a different prediction for the long-term agent density in a growth-to-confluence experiment~\cite{tremel2009cell}. This approach was subsequently extended to derive an approximate PDE for agent density in a non-uniform, invading population~\cite{plank2013lattice}, and a population with directionally biased movement~\cite{irons2015lattice}. The low-density behaviour at the travelling wavefront was accurately captured by both the approximate PDE derived from the lattice-free model and the Fisher-KPP equation (which is the continuum-limit of an equivalent lattice-based model). However, only the lattice-free approach captured the high-density behaviour behind the wavefront, where agent interactions become more important.


\section{Other types of interactions and model extensions} 
\label{sec:other_interactions}

In Section~\ref{sec:crowding}, we considered models of crowding effects where agents interact by competing locally for space. In this section, we consider other types of agent-agent interaction that can arise, together with associated extensions to the basic modelling framework. 


\subsection{Lattice-based models of local adhesion and repulsion} 
\label{sec:adhesion_repulsion}

In addition to considering a lattice-based exclusion process to describe biased or unbiased motion, we can also introduce additional interactions such as adhesion (attraction) or repulsion between neighbouring agents on the lattice~\cite{deroulers2009modeling,Schmeiser2009,fernando2010nonlinear,johnston2012mean,Thompson2012}. These kinds of mechanisms can be motivated by considering experimental images of the spreading of cancer cells which appear to clump together, thereby motivating the introduction of an adhesive mechanism.

Box~3 illustrates how local adhesion or repulsion between neighbouring agents can be included in a lattice-based exclusion process model. In this model, isolated agents undergo an unbiased random walk, but agents that have a neighbouring agent at an adjacent lattice site have a movement probability that is modified via an additional parameter $\sigma\in[-1, 1]$. Setting $\sigma > 0$ models adhesion by reducing the probability of moving away from a neighbouring agent, while $\sigma < 0$ models repulsion by increasing the probability of moving away from a neighbouring agent~\cite{Schmeiser2009,johnston2012mean}.



\bigskip
\begin{tcolorbox}[breakable, title=Box 3: Lattice-based random walk model and continuum-limit PDE with exclusion and adhesion/repulsion]
\parskip=6pt

To model adhesion/repulsion between neighbouring agents, suppose that the probability that an agent moves from site $(i,j)$ to site $(i+1,j)$ is defined by
\begin{equation} \label{eq:movement_probability_adhesion}
P_{(i,j)\to(i+1,j)}  = \frac{P_m}{4}U_{i,j} (1-\sigma U_{i-1,j})(1-U_{i+1,j}),
\end{equation}
with similar formulations for the probability of movement to $(i-1,j)$ and $(i,j\pm 1)$.
The factor of $(1-U_{i+1,j})$ in Equation~\eqref{eq:movement_probability_adhesion}  models volume exclusion in the same way as in Box~2. The factor of $(1-\sigma U_{i-1,j})$ in Equation~\eqref{eq:movement_probability_adhesion} is a simple model of adhesion/repulsion with the parameter $\sigma \in [-1,1]$ controlling the strength and nature of the effect. Setting $\sigma > 0$ decreases the probability of moving away from a neighbouring agent, modelling adhesion, whereas setting $\sigma < 0$ increases this probability, modelling repulsion. 

Making the assumption that the occupancy of adjacent lattice sites is independent, the change in the average occupancy $u_{i,j}$ of site $(i,j)$ during a time step $\tau$ may be approximated as: 
\begin{align}
\Delta u_{i,j}  = & \,\, \dfrac{P_m}{4}  \underbrace{\left(1 - u_{i,j}\right)\left[u_{i-1,j}(1 - \sigma u_{i-2,j})+u_{i+1,j}(1 - \sigma u_{i+2,j})  \right]  }_{\textrm{migration onto site $(i,j)$}} \label{eq:conservation_eqn_adhesion} \\
&+ \dfrac{P_m}{4}  \underbrace{\left(1 - u_{i,j}\right)\left[u_{i,j-1}(1 - \sigma u_{i,j-2})+u_{i,j+1}(1 - \sigma u_{i,j+2})  \right]  }_{\textrm{migration onto site $(i,j)$}} \notag \\
&-\dfrac{P_m}{4}\underbrace{u_{i,j} \left[(1-\sigma u_{i-1,j})(1-u_{i+1,j}) + (1-\sigma u_{i+1,j})(1-u_{i-1,j})   \right]}_{\textrm{migration out of site $(i,j)$}} \notag  \\
&-\dfrac{P_m}{4}\underbrace{u_{i,j} \left[(1-\sigma u_{i,j-1})(1-u_{i,j+1}) + (1-\sigma u_{i,j+1})(1-u_{i,j-1})   \right]}_{\textrm{migration out of site $(i,j)$}}. \notag  
\end{align}
Again, we identify $u_{i,j}$ with a smooth function $u(x,y,t)$ and expand all terms in Equation~\eqref{eq:conservation_eqn_adhesion} as Taylor series about $(x,y) = (i\delta, j\delta)$. Dividing the resulting expressions by $\tau$ and taking the limit as $\delta \to 0$ and $\tau \to 0$ with $\delta^2/\tau$ held constant gives
\begin{equation} \label{eq:PDE_adhesion}
\dfrac{\partial u}{\partial t} =  \dfrac{\partial }{\partial x}\left[D(u)\dfrac{\partial u}{\partial x}\right] + \dfrac{\partial }{\partial y}\left[D(u)\dfrac{\partial u}{\partial y}\right],  
\end{equation}
where
\begin{equation} \label{eq:diffusivity_with_adhesion}
D(u) = D_0\left( 1 - \sigma  u(4-3u) \right), \qquad D_0  = \lim_{\substack{\delta \to 0 \\ \tau \to 0}} \left( \dfrac{P_m \delta^2}{4 \tau} \right). 
\end{equation}

\end{tcolorbox}



In this model the continuum-limit equation is a nonlinear diffusion equation, Equation \eqref{eq:PDE_adhesion}, with density-dependent diffusivity, $D(u)$, given by Equation~\eqref{eq:diffusivity_with_adhesion}. If $\sigma = 0$ the diffusivity relaxes to $D(u)=D_0$ and we recover the linear diffusion equation.  If $\sigma > 0$ we have $D(u)  < D_0$, which means spreading of the population is slower than in the $\sigma=0$ case, and vice versa if $\sigma<0$.  The magnitude of the effect on the diffusivity is strongest in high-density regions ($u\approx 1$), where agent interactions occur frequently, and minimal in low-density regions ($u\ll 1$), where agent interactions are rare. 

To illustrate the behaviour of this model, we consider an ensemble of simulations with same lattice and initial condition as previously, and $\sigma = 3/4$ or $\sigma = -3/4$ (Figure~\ref{fig:pde_plots}e--f). Comparing results for different choices of $\sigma$ confirms that setting $\sigma > 0$ to model adhesion slows the outward spread of the initially-confined population  (Figure~\ref{fig:pde_plots}e) relative to the case with $\sigma = 0$, which represents an unbiased exclusion process with no adhesion/repulsion (Figure~\ref{fig:pde_plots}c). Setting $\sigma < 0$ to model repulsion increases the outward spread of the population (Figure~\ref{fig:pde_plots}f). For this choice of parameters the numerical solution of the nonlinear diffusion equation accurately predicts the average behaviour of the stochastic simulations.  Similar ideas have been used to model migration on a lattice with bias, adhesion and volume exclusion~\cite{SimpsonMcElwainUpton2010}, and have been applied in the lattice-free setting~\cite{Johnston2013}.  

One point to note is that by rewriting the diffusivity in Equation~\eqref{eq:diffusivity_with_adhesion} as $D(u)=D_0[3\sigma(u-2/3)^2+1-4\sigma/3]$, we see that setting $3/4 < \sigma < 1$ leads to a situation where $D(u) < 0$ for a range of values of $u$ centered around $u=2/3$~\cite{Schmeiser2009}. In this case of strong adhesion, stochastic simulation results can lead to visually obvious clustering of agents~\cite{Simpsonmeso}.  While it is possible to numerically solve the continuum-limit PDE model when $D(u) < 0$ for a range of $u$, giving rise to shock-fronted solutions, these solutions may no longer provide a good match to averaged data from the ABM in these extreme cases~\cite{Schmeiser2009,johnston2012mean}.

\subsection{Multi-type models}

Inspired by different types of biological observations and experiments, there have been many extensions of lattice-based and lattice-free ABMs. All of the models discussed so far involve simulating the random motion of a populations of agents that all behave in the same way and this is appropriate for modelling a population of functionally identical individuals. However, many biological experiments and ecological observations include individuals from multiple populations (e.g.~cancer cells moving within a population of skin cells, or predators interacting with prey). These kinds of scenarios have also been modelled using lattice-based approaches, for example such models have been used to describe donor-host transplant cell biology experiments~\cite{Simpson2009}, where agents can be of different types to reflect the different species in the system.

A related model extension is to subdivide a population of agents according to some important biological characteristic. For example, models have been used to simulate the progression of cells in a population through the cell cycle, as revealed by fluorescent cell-cycle labels~\cite{GAVAGNIN201991,SIMPSON2018}. Similar ideas have been used to model other kinds of life histories, such as explicitly modelling adult and juvenile sub-populations, differentiation of stem cells~\cite{reina2012lattice}, or tracking the age or size of individual agents~\cite{adams2013growth}. 

This framework enables various types of interactions to be included in lattice-based models, such as transition of agents from one type to another~\cite{SIMPSON2018}, predation~\cite{molina2015analyzing}, interspecific competition~\cite{durrett1998spatial,ying2014species}, Allee effects~\cite{windus2007allee}, disease transmission~\cite{rhodes1997epidemic,keeling1999effects}, or the action of an anti-mitotic drug~\cite{jin2021mathematical}. Taking the continuum limit of a multi-type ABM typically leads to a system of PDEs for the density of each type of agent, which are coupled via reaction terms representing transitions or interactions between agents of different types~\cite{Simpson2009,bruna2012diffusion}. 


\subsection{Modelling domain growth}

Another important generalisation of lattice-based ABMs arises in the field of developmental biology, where the random motion of molecules and cells during embryonic development can be influenced by the growth of those tissues. These processes have been modelled using both non-interacting lattice-based models~\cite{Baker2010,Yates2012,Yates2013} and lattice-based exclusion process models~\cite{Binder2008}. A key challenge in lattice-based models of random motility on a growing domain is to determine how changes in domain length, which is typically simulated via random insertion or deletion of lattice sites~\cite{YATES2014}, affects the distribution of agents on the lattice. Taking the continuum limit of lattice-based models that include domain growth typically leads to additional advection terms that describe how growth affects the spatiotemporal distribution of agent density~\cite{Baker2010,Simpson2015}. 


\section{Accounting for spatial structure: going beyond the mean-field} 
\label{sec:spatial_structure}

In most of the approaches to modelling agent interactions discussed in Sections \ref{sec:crowding} and \ref{sec:other_interactions}, the continuum-limit approximation uses the mean-field assumption, which is that the probability of an agent being at location $\vect{x}$ is unaffected by the presence of an agent at another location $\vect{x'}$. This simplifies the analysis, but ignores short-range spatial correlations that may arise as a result of crowding effects, dispersal and other biophysical processes. As we shall see in this section, such correlations can also impact the dynamics at the population-level, because the extent to which agents are clustered together or spaced apart affects the frequency and strength of agent interactions.

A patchy distribution of agents can arise from a number of different mechanisms, for example, through environmental heterogeneity or other external factors, or solely through interactions among agents. Although these different mechanisms may be difficult to distinguish by looking at the distribution of agents in a single model realisation, they will generally be distinguishable given data from multiple realisations. Environmental heterogeneity and other external factors will typically drive the formation of patches in the same spatial locations, whereas patterns of agent density generated solely from agent-agent interactions will have predictable statistics but the locations of individual patches will differ from one realisation to the next. In this section, we are primarily concerned with this latter case, although the two are not mutually exclusive and can operate in combination.

One way to account for correlations in agent locations is to use a {\em spatial moments} approach, which tracks not only the mean agent density (i.e.~the first spatial moment), but also the mean density of pairs of agents (i.e.~the second spatial moment). This approach has its origins in statistical physics~\cite{kirkwood1935statistical} but has increasingly found applications in biology -- see, for example,~\cite{matsuda1992statistical,bolker1997using,keeling2000multiplicative,lewis2000modeling}. A drawback of this approach is that a closure approximation is usually needed to obtain a tractable system~\cite{dieckmann2000relaxation,bolker2003combining}. This typically involves approximating the third moment in terms of the first and second moments and the choice of approximation is not unique~\cite{murrell2004moment,raghib2011multiscale}. However, it is important to recognise that mean-field models implicitly close the system at the level of the first moment, and that by closing at the level of the second moment some information about the effects of spatial structure is retained. In Sections \ref{sec:spatial_structure_lattice_based}--\ref{sec:spatial_moments_applications}, we describe how the spatial moments approach has been used for different types of ABM.


\subsection{Lattice-based models} 
\label{sec:spatial_structure_lattice_based}

In deriving the mean-field limit for lattice-based exclusion processes in Section~\ref{sec:crowding}, for example in Box~2, we assumed that the random variables representing the occupancy status of two different lattice sites $U_{i,j}$ and $U_{k,l}$ were independent, which implies that $\langle U_{i,j} U_{k,l}\rangle = u_{i,j} u_{k,l}$. As a consequence of this mean-field assumption, expressions appearing in the conservation statement in Equation~\eqref{eq:conservation_eqn_exclusion} can be simplified. For example, the probability that there is an agent at $(i,j)$ and that agent moves to $(i\pm 1,j)$ is
\begin{equation} \label{eq:example_transition_probability}
P_{(i,j)\to(i\pm1,j)}=\frac{P_m}{4} \langle U_{i,j}(1-U_{i\pm 1,j})\rangle.
\end{equation}
Applying the mean-field assumption and using linearity of expectations, this may be approximated as $(P_m/4) u_{i,j}(1-u_{i\pm 1,j})$, where $u_{i,j}$ denotes $\langle U_{i,j}\rangle$ as previously.

Including correlations in the agent locations recognises that occupancy statuses of different lattice sites are not generally independent, at least when the distance between sites is relatively small. Box~4 provides a worked example for the lattice-based exclusion process model with proliferation considered in Section~\ref{sec:crowding_lattice_based_with_proliferation}.

\bigskip
\begin{tcolorbox}[breakable,title=Box 4. Effect of pair correlations in a lattice-based model]
\parskip=6pt

Consider a two-dimensional lattice-based model where, during each time step $\tau$, agents attempt to move with probability $P_m$ and attempt to proliferate with probability $P_p$. Suppose that the target site for all attempted movement and proliferation events is randomly selected from the four nearest-neighbour sites (i.e.~movement is unbiased) and any event where the target site is occupied is aborted. 

To incorporate the joint dependence on the occupancy status of the starting site and the target site into the expression for  $P_{(i,j)\to(i\pm1,j)}$ in Equation~\eqref{eq:example_transition_probability}, we use $u_2(1(i,j),0(i\pm 1,j))$ to denote the joint probability that site $(i,j)$ is occupied by an agent and site $(i\pm1,j)$ is empty. Using this notation, Equation~\eqref{eq:example_transition_probability} may be written
\begin{equation}
P_{(i,j)\to(i\pm1,j)}=\frac{P_m}{4}u_2(1(i,j),0(i\pm1,j)).
\end{equation}
Then, the change in the average occupancy $u_{i,j}$ of site $(i,j)$ during a time step $\tau$ is  
\begin{align}
\Delta u_{i,j}  = & \,\, \dfrac{P_m}{4} \underbrace{\left[u_2(0(i,j),1(i-1,j)+u_2(0(i,j),1(i+1,j)\right]}_{\textrm{migration onto site $(i,j)$}} \label{eq:conservation_eqn_moments} \\
&+\dfrac{P_m}{4} \underbrace{\left[u_2(0(i,j),1(i,j-1)+u_2(0(i,j),1(i,j+1) \right]  }_{\textrm{migration onto site $(i,j)$}} \notag \\
&-\dfrac{P_m}{4}\underbrace{\left[u_2(1(i,j),0(i-1,j)+u_2(1(i,j),0(i+1,j)\right]}_{\textrm{migration out of site $(i,j)$}} \notag  \\
&-\dfrac{P_m}{4}\underbrace{\left[u_2(1(i,j),0(i,j-1)+u_2(1(i,j),0(i,j+1)\right]}_{\textrm{migration out of site $(i,j)$}} \notag \\
&+\dfrac{P_p}{4}\underbrace{\left[u_2(0(i,j),1(i-1,j)+u_2(0(i,j),1(i+1,j)\right]}_{\textrm{proliferation onto site $(i,j)$}}\notag\\
&+\dfrac{P_p}{4}\underbrace{\left[u_2(0(i,j),1(i,j-1)+u_2(0(i,j),1(i,j+1)\right]}_{\textrm{proliferation onto site $(i,j)$}}.\notag
\end{align}
To make progress in deriving a continuum-limit equation we define the correlation in occupancy probability between two neighbouring lattice sites, $F$, via
\begin{equation} \label{eq:pair_density_lattice}
u_2(1(i,j),1(i\pm1,j\pm1)=u_{i,j}u_{i\pm1,j\pm1}F((i,j),(i\pm1,j\pm1)).
\end{equation}
Using conservation of total probability we can simplify some terms in Equation~\eqref{eq:conservation_eqn_moments} by noting, for example, that
\begin{equation}
u_2(1(i,j),1(i\pm1,j\pm1)+u_2(1(i,j),0(i\pm1,j\pm1)=u_{i,j}.
\end{equation}
These ideas enable us to simplify Equation~\eqref{eq:conservation_eqn_moments} to give
\begin{align}
\Delta u_{i,j}  = & \,\, \dfrac{P_m}{4}\underbrace{\left(1-u_{i,j}\right)\left[u_{i-1,j}+ u_{i+1,j}+ u_{i,j-1}+u_{i,j+1}\right] }_{\textrm{migration onto site $(i,j)$}} \label{eq:conservation_eqn_momentsF} \\
&-\dfrac{P_m}{4}\underbrace{u_{i,j}\left[\left(1-u_{i+1,j}\right)+\left(1-u_{i-1,j}\right) +\left(1-u_{i,j-1}\right) +\left(1-u_{i,j+1}\right)  \right]}_{\textrm{migration out of site $(i,j)$}} \notag \\ 
&+\dfrac{P_p}{4}\underbrace{u_{i-1,j}(1-F((i,j),(i-1,j))u_{i,j})+u_{i+1,j}(1-F((i,j)(i+1,j))u_{i,j})}_{\textrm{proliferation onto site $(i,j)$}}\notag\\
&+\dfrac{P_p}{4}\underbrace{u_{i,j-1}(1-F((i,j),(i,j-1)u_{i,j})+u_{i,j+1}(1-F((i,j)(i,j+1)u_{i,j})}_{\textrm{proliferation onto site $(i,j)$}}.\notag
\end{align}
As in the mean-field case, we identify $u_{i,j}$ with a smooth function $u(x,y,t)$ and expand all terms in Equation~\eqref{eq:conservation_eqn_momentsF} as Taylor series about $(x,y) = (i\delta, j\delta)$. Dividing the resulting expressions by $\tau$ and taking the limit as $\delta \to 0$ and $\tau \to 0$ with $\delta^2/\tau$ held constant gives
\begin{equation}
\dfrac{\partial u}{\partial t} =  \underbrace{D\left[\dfrac{\partial^2u}{\partial x^2} + \dfrac{\partial^2u}{\partial y^2} \right]}_{\textrm{random motility}} + \underbrace{\vphantom{\left[\dfrac{\partial^2u}{\partial x^2}\right]}ru(1-f(x,y,t)u),}_{\textrm{proliferation}} 
\label{eq:PDE_moments}
\end{equation}
where
\begin{equation}
D  = \lim_{\substack{\delta \to 0 \\ \tau \to 0}} \left( \dfrac{P_m \delta^2}{4 \tau} \right), \quad 
r  = \lim_{\substack{\tau \to 0}} \left( \dfrac{P_p}{\tau} \right).
\end{equation}
We interpret $f(x,y,t)$ a function that accounts for the influence of local density correlations upon the proliferation rate at time $t$. 

\end{tcolorbox}

The continuum-limit equation for mean agent density, Equation~\eqref{eq:PDE_moments}, contains a linear diffusion term and a proliferation term containing a function $f$ that quantifies the impact of local correlations on population growth. The mean-field model assumes that $f(x,y,t)\equiv1$, which gives rise to the Fisher-KPP equation describing the growth and spread of the population. However, in reality $f(x,y,t)\geq1$ for this proliferation and migration model, which means that local correlations (arising through the proliferation-driven build up of clusters of agents on the lattice) suppress the growth rate of the population. Typically, larger values of $P_p/P_m$ entail larger values for $f(x,y,t)$~\cite{baker2010correcting}. Equation~\eqref{eq:PDE_moments} also highlights that for unbiased motility and no proliferation, the mean-field assumption is exact: all terms containing the correlations cancel~\cite{simpson2011corrected}. This is not the case, however, when movement is biased.

To make progress in deriving an expression for $f(x,y,t)$ we need to derive equations for the rate of change of the correlation functions $F((i,j),(i\pm1,j\pm1))$ in Equation~\eqref{eq:conservation_eqn_momentsF}. These equations will depend on both the correlation in occupancy probabilities of sites that are ``next nearest neighbours'' as well as on the joint occupancy probabilities of triples of lattice sites~\cite{baker2010correcting}. Closing this system of ``correlation equations'' requires deriving the system of equations for the evolution of all of the pairwise correlation functions $F((i,j),(k,l))$, for all pairs of sites $(i,j)$ and $(k,l)$ on the lattice, and then closing this system by using a closure approximation for the triplet occupancy probability terms~\cite{kirkwood1935statistical,matsuda1992statistical,murrell2004moment}. In general, the method is unwieldy, but it can produce significantly improved predictions of the evolution of agent density over time, in particular in two and three spatial dimensions~\cite{baker2010correcting}.

As with mean-field models, different types of behaviours and interactions can be included. In the cell biology literature, it is typical to include cell death at a constant rate $\mu$, independent of local cell density. This results in an additional term of the form $-\mu u$ in Equation~\eqref{eq:PDE_moments}. In this case, the death terms tend to reduce local correlations whilst the proliferation terms increase them.
The framework has been extended to include multiple species~\cite{Markham2013,JOHNSTON201581}, multi-scale interactions \cite{ellner2001pair}, domain growth~\cite{ROSS2015,Ross2016}, more general types of ecological interactions~\cite{iwasa2000}, and coexistence of competing species~\cite{ying2014species}.


\subsection{Lattice-free models} 
\label{sec:spatial_structure_lattice_free}

Spatial correlations can be also incorporated into lattice-free models, using analogous methods to the lattice-based approach in Section~\ref{sec:spatial_structure_lattice_based}. The locations of a population of agents in continuous space at a given time define a {\em spatial point process}, which may be described by its {\em spatial moments} (also known as {\em product densities})~\cite{illian2008statistical,finkelshtein2009individual}. The $k^\mathrm{th}$ spatial moment $p^{(k)}$ is defined such that $p^{(k)}(\vect{x}_1,\ldots,\vect{x}_k)\text{d}V_1\ldots\text{d}V_k$ is the probability that there is an agent in each of the $k$ infinitesimal volumes $\text{d}V_1,\ldots,\text{d}V_k$ around locations $\vect{x}_1,\ldots,\vect{x}_k$. Thus, the first moment $p^{(1)}(\vect{x})$ is just the mean agent density at location $\vect{x}$. The second moment $p^{(2)}(\vect{x},\vect{x}')$ is the density of pairs of agents at locations $\vect{x}$ and $\vect{x}'$ and so on. Note that $p^{(2)}(\vect{x},\vect{x}')$ is analogous to the pairwise lattice site occupancy probability in Equation~\eqref{eq:pair_density_lattice}.

When the population size and/or locations of agents changes over time, the spatial moments are time-dependent. The dynamics of the spatial point process may be represented by a set of equations for the rate of change of each spatial moment. The simplest case is where there are no interactions between agents, which means that the dynamics of the $k^\mathrm{th}$ moment do not depend on higher-order moments, so the system is closed at the level of the first (or any higher-order) moment. Here, movement and/or dispersal processes may be defined by {\em kernels} representing the probability density function (usually in two spatial dimensions) for the movement vector or for the dispersal vector of the daughter agent from the parent. This naturally gives rise to an integro-differential equation (or integro-difference equation in discrete time), in which the change in mean density at $\vect{x}$ is calculated by integrating over the density at neighbouring locations~\cite{kot1996dispersal,hastings2005spatial}. For example, if the probability of an agent at location $\vect{x}$ moving to location $\vect{x}'$ during a small time step $\tau$ is $\lambda k_m(\vect{x},\vect{x}') \tau$ and all agents move independently, then the dynamics of mean agent density are described by
\begin{equation} \label{eq:integro_diff}
\frac{\partial p^{(1)}(\vect{x})}{\partial t} = \underbrace{-\lambda p^{(1)}(\vect{x}) \vphantom{\int} }_{\textrm{migration away from } \vect{x}} + \underbrace{\lambda\int p^{(1)}(\vect{x}') k_m(\vect{x}',\vect{x})\,\text{d}\vect{x}'}_{\textrm{migration to } \vect{x}} .
\end{equation}
The first term in Equation~\eqref{eq:integro_diff} represents agents moving away from location $\vect{x}$. The second term represents agents moving to location $\vect{x}$ by integrating the density over all possible starting locations $\vect{x}'$, weighted by the probability $k_m(\vect{x}',\vect{x})$ of moving from $\vect{x}'$ to $\vect{x}$. It is common for the movement kernel $k_m$ to depend only on the relative displacement, $\vect{\xi}=\vect{x}-\vect{x}'$, and so to be independent of the starting location $\vect{x}$, which simplifies Equation~\eqref{eq:integro_diff}. In this case and when movement is unbiased, it is worth noting that in the limit where the variance $\sigma^2$ of the movement kernel $k_m(\vect{\xi})$ tends to zero such that $ \lambda \sigma^2$ is held constant, Equation~\eqref{eq:integro_diff} reduces to the linear diffusion equation with diffusivity $D=\lim_{\lambda,\delta\to 0} (\lambda \delta^2/4)$. This is equivalent to the diffusivity in Equation~\eqref{eq:diffusivity_drift} for a lattice-based random walk as $\lambda$ is equivalent to the probability of movement per unit time, $P_m/\tau$.

Interactions between neighbouring agents that lead to local density-dependence in the rates of proliferation, death, or movement may also be described by kernel functions $k(\vect{\xi})$ representing the relative contribution of a neighbour located at $\vect{x}+\vect{\xi}$ to the probability of a specific event for an agent located at $\vect{x}$. Pairwise interactions of this kind generally induce a dependence of the rate of change of the $k^\mathrm{th}$ moment on the $(k+1)^\mathrm{th}$ moment, and this leads to an infinite hierarchy of equations for $\partial p^{(k)}/\partial t$ ($k=1,2,\ldots$). To obtain a closed system, one approach is to approximate the $(k+1)^\mathrm{th}$ moment in terms of lower-order moments using a moment closure approximation~\cite{murrell2004moment}. 

An important special case is where the point process is {\em spatially stationary}, meaning that the first moment is independent of location $\vect{x}$ and the second moment depends only on the relative locations of the agents in the pair, $\vect{\xi}=\vect{x}'-\vect{x}$~\cite{illian2008statistical}.
In this case, the ratio $g(\vect{\xi}) = p^{(2)}(\vect{\xi})/[p^{(1)}]^2$, referred to as the {\em pair correlation function}, provides information about the type of spatial structure~\cite{illian2008statistical}, analogous to the lattice-based correlation function $F$ in Equation~\eqref{eq:conservation_eqn_momentsF}. If $g(\vect{\xi})=1$, there are no pairwise correlations in agent locations and the spatial structure (up to the level of second moment) is completely random, as for a Poisson point process. If $g(\vect{\xi})<1$ for $|\vect{\xi}|<l$, then a neighbourhood of radius $l$ around an agent is likely to contain fewer agents than in a Poisson point process. This is referred to as {\em regular} spatial structure. If $g(\vect{\xi})>1$ for $|\vect{\xi}|<l$, the neighbourhood is likely to contain more agents than in a Poisson point process. This is referred to as {\em clustered} spatial structure. Box~5 provides a worked example for the {\em spatial logistic model}~\cite{bolker1997using,law2003population}, a lattice-free version of the lattice-based logistic model~\cite{matsuda1992statistical,ellner2001pair}. 



\begin{figure}
	\includegraphics[width=\textwidth, trim={1cm 7cm 1cm 3cm},clip]{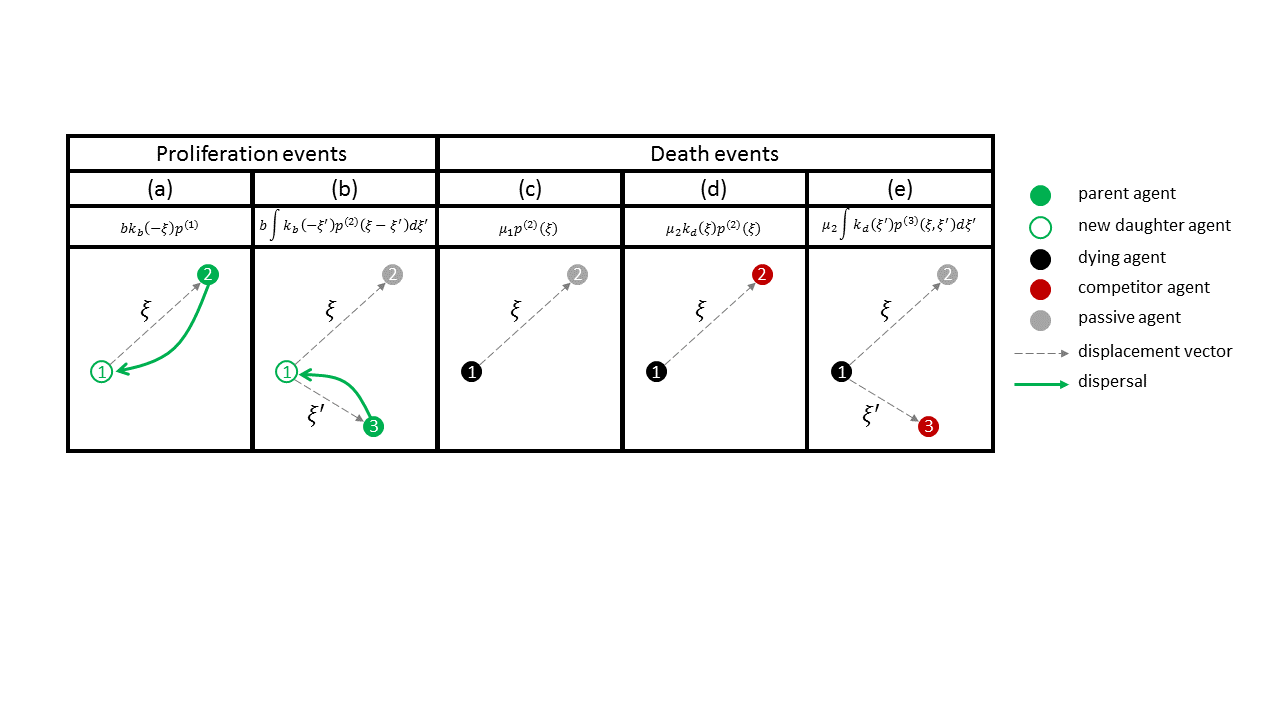}
	\caption{Spatial logistic model: illustration of the types of events that create or destroy a pair of agents separated by displacement vector $\vect{\xi}$ (see Box~5). (a) proliferation of a single agent and dispersal of the daughter by $-\vect{\xi}$; (b) proliferation of an agent in a pair with a passive neighbour at $\vect{\xi}-\vect{\xi}'$ and dispersal of the daughter by $-\vect{\xi}'$; (c) density-independent death of an agent in a pair with a passive neighbour at $\vect{\xi}$; (d) neighbour-dependent death of an agent caused by the other agent in the pair at $\vect{\xi}$; (e) neighbour-dependent death of an agent in a pair with a passive neighbour at $\vect{\xi}$ caused by a competitor at $\vect{\xi}'$. The expressions above the diagrams show the rate at which each event type occurs.  }
	\label{fig:spatial_moments}
\end{figure}

\bigskip
\begin{tcolorbox}[breakable,title=Box 5. The spatial logistic model]
\parskip=6pt

Consider a population of agents undergoing proliferation and death with no movement where $\vect{x}_i$ denotes the location of agent $i$. Suppose that the proliferation rate $b$ is independent of the local density and daughters of agent $i$ are dispersed to a location $\vect{x}_i+\vect{\xi}$ with probability density function $k_b(\vect{\xi})$. Suppose also that the death rate has both a density-independent component, $\mu_1$, and a component that depends on the density in the local neighbourhood, $\mu_2 \sum_j k_d(\vect{x}_j-\vect{x}_i)$. This represents the effects of local competition among neighbours.

In this example, we assume that the system is translationally invariant, i.e.~the point process representing the agent locations is spatially stationary. This means that the first spatial moment is independent of $\vect{x}$ and the second spatial moment depends only on the displacement $\vect{\xi}$ between the agents in the pair~\cite{bolker1997using,bolker1999spatial}. The rate of change of the first moment is equal to expected proliferation minus expected deaths:
\begin{equation} \label{eq:1st_moment}
	\frac{\text{d}p^{(1)}}{\text{d}t} = \underbrace{(b-\mu_1) p^{(1)} \vphantom{\int}}_\textrm{density-independent proliferation-death} - \underbrace{\mu_2 \int k_d(\vect{\xi}) p^{(2)}(\vect{\xi}) \,\text{d}\vect{\xi}.}_\textrm{neighbour-dependent death}
\end{equation}
The integral term in Equation~\eqref{eq:1st_moment} represents the expected density of pairs of agents $p^{(2)}(\vect{\xi})$ separated by displacement vector $\vect{\xi}$, weighted by the influence $k_d(\vect{\xi})$ that a neighbour at displacement $\vect{\xi}$ has on the death rate. 
The rate of change of the second moment is given by the balance between processes that create or destroy a pair of agents separated by displacement $\vect{\xi}$: 
\begin{eqnarray} 
\label{eq:2nd_moment}
	\frac{\partial p^{(2)}(\vect{\xi})}{\partial t} &=&\underbrace{bk_b(-\vect{\xi})p^{(1)} \vphantom{\int}}_\mathrm{a} + \underbrace{b \int k_b(-\vect{\xi}') p^{(2)}(\vect{\xi}-\vect{\xi}') \,\text{d}\vect{\xi}'}_\mathrm{b}      \\
	&& - \underbrace{\mu_1 p^{(2)}(\vect{\xi}) \vphantom{\int}}_\mathrm{c} 
	 -\underbrace{\mu_2 k_d(\vect{\xi})p^{(2)}(\vect{\xi})\vphantom{\int} }_\mathrm{d}
	 - \underbrace{\mu_2 \int k_d(\vect{\xi}') p^{(3)}(\vect{\xi},\vect{\xi}') \,\text{d}\vect{\xi}'}_\mathrm{e}. \notag
\end{eqnarray}
Figure~\ref{fig:spatial_moments} illustrates the events that are represented by terms a-e: each term represents a type of event that creates or removes the agent at location $1$ in Figure~\ref{fig:spatial_moments}. For simplicity, symmetric terms for the creation or removal of the agent at location $2$ are omitted, see~\cite{plank2015spatial}. 

If we make the mean-field assumption that the mean pair density is equal to the square of the mean agent density (i.e.~$p^{(2)}(\vect{\xi})=[p^{(1)}]^2$) then, since $k_d(\vect{\xi})$ is a probability density function, Equation~\eqref{eq:1st_moment} reduces to the logistic growth ODE:
\begin{equation} 
\label{eq:logsitc_ODE}
\frac{\text{d}p^{(1)}}{\text{d}t} = (b-\mu_1)p^{(1)} - \mu_2 [p^{(1)}]^2. 
\end{equation}
Thus, models that use a logistic growth term to represent proliferation with local dispersal and local competition among neighbours implicitly make a moment closure assumption at the level of the first moment. 

\end{tcolorbox}

\bigskip


Some information about the spatial structure in the ABM is retained by working with Equation~\eqref{eq:2nd_moment} and closing the system at the level of the second moment via a closure approximation for $p^{(3)}(\vect{\xi},\vect{\xi}')$~\cite{murrell2004moment}. This extra information not only enables quantification of spatial patterns in the population, it can crucially affect the dynamics of mean agent density via the second term in Equation~\eqref{eq:1st_moment}. Closing the system at level of the second moment ($k=2$) has been found to provide a reasonable approximation to the underlying ABM in many cases where the mean-field approximation breaks down. 

\begin{figure}
    \centering
    \includegraphics[width=0.68\textwidth, trim={1.6cm 1.5cm 1.6cm 1cm},clip]{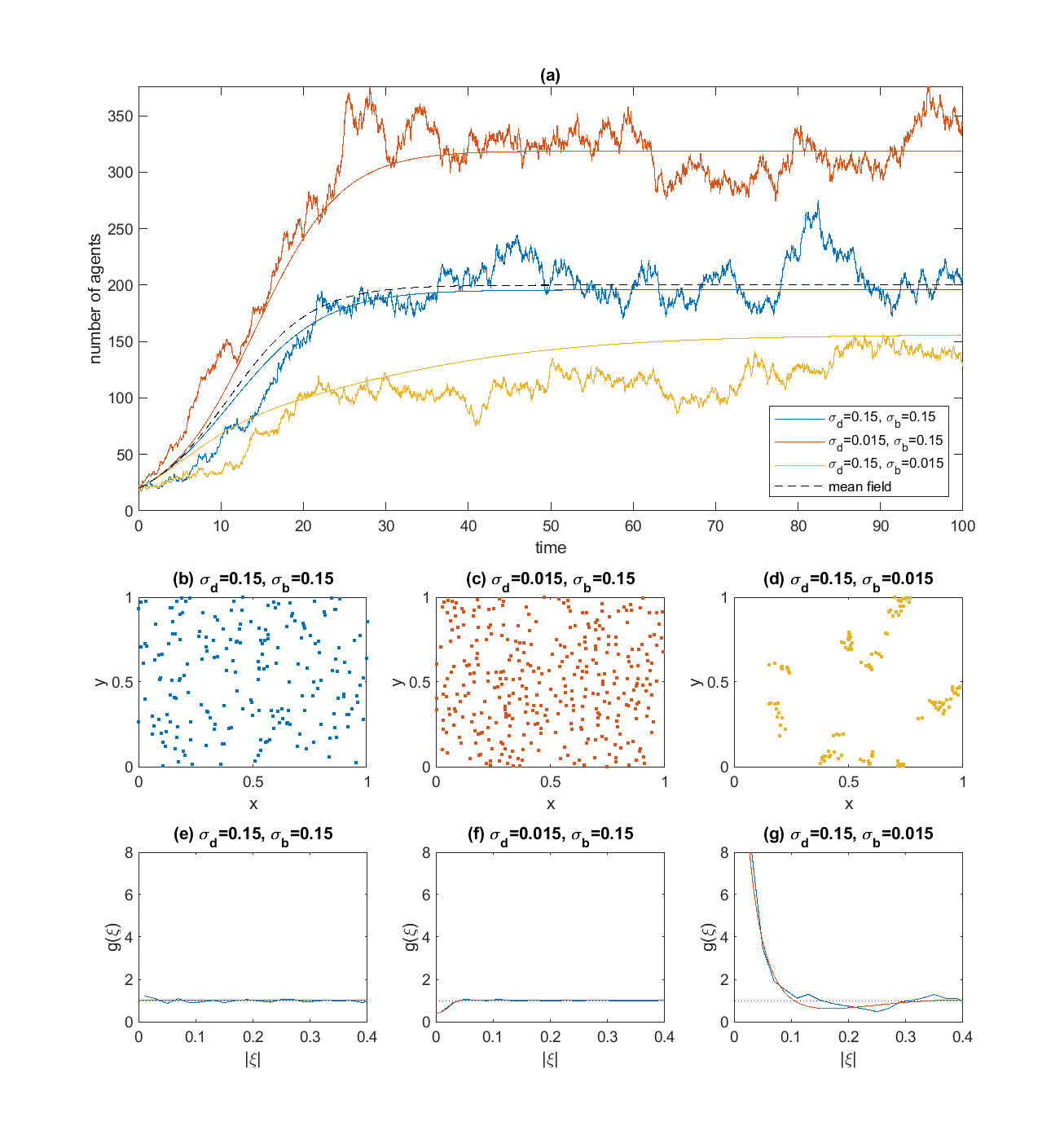}
    \caption{Spatial logistic model. (a) Population size as a function of time for long-range competition and dispersal (blue), short-range competition and long-range dispersal (red) and long-range competition and short-range dispersal (yellow). Noisy, coloured curves show a single simulation of the ABM; smooth, coloured curves show the numerical solution of the spatial moment dynamics model closed at the level of the second moment via Equation~\eqref{eq:2nd_moment} with a 4-1-1 power-2 closure; dashed black curve shows the solution of the mean-field model, Equation~\eqref{eq:logsitc_ODE}. (b)--(d) show agent locations at $t=100$. (e)--(g) show the pair correlation function $g(\vect{\xi})$ calculated from the ABM simulation (blue) and the spatial moment dynamics model (red). In (b) and (e) the spatial structure is approximately Poisson (i.e.~$g(\vect{\xi})\approx 1$) and population size is close to that predicted by the mean-field model. In (c) and (f) the spatial structure is regular (i.e.~$g(\vect{\xi})<1$ for small $|\vect{\xi}|$) and the population size is higher than predicted by the mean-field model. In (d) and (g) the spatial structure is clustered (i.e.~$g(\vect{\xi})>1$ for small $|\vect{\xi}|$) and the population size is lower than that predicted by the mean-field model. Competition and dispersal kernels $k_d$ and $k_b$ are Gaussian functions with standard deviation $\sigma_d$ and $\sigma_b$, respectively. The model was initialised with $20$ agents uniformly distributed in the domain $[0,1]\times[0, 1]$ with periodic boundary conditions. Other parameter values are $b=0.4$, $\mu_1=0.2$, $\mu_2=0.001$. See Box~5 and~\cite{law2000dynamical,law2003population} for details.}
    \label{fig:slm} 
\end{figure}

For example, Figure~\ref{fig:slm} shows example simulations of the spatial logistic model for different values of the parameters representing the length scale over which agents compete with their neighbours and the length scale for dispersal of daughter agents. Note that, in this example, the model is isotropic, meaning that the kernels $k(\vect{\xi})$ and the second spatial moment $p^{(2)}(\vect{\xi},t)$ at any given time $t$ depend only on $|\vect{\xi}|$ and not on the direction of $\vect{\xi}$. Short-range competition leads to a regular spatial structure. This reduces the overall death rate (because the average density in the neighbourhood of an agent is lower than the average density overall) and thus allows the population to reach a higher density than the mean-field model predicts (Figure~\ref{fig:slm}a,c,f). Conversely, short-range dispersal leads to a clustered spatial structure, which elevates the overall death rate and reduces the average density below that predicted by the mean-field model (Figure~\ref{fig:slm}a,d,g)~\cite{law2003population}). Similar results have been found in a comparable lattice-based model \cite{ellner2001pair}. 


More generally, short-range dispersal, adhesion or attraction between agents will tend to generate clustered structure. Short-range competition or crowding effects will tend to generate regular structure. Random motility and neighbour-independent death will tend to weaken any spatial structure. These processes can operate at different spatial scales and can impact the population-level dynamics, as well as the fine-scale distribution of agents. 

The accuracy of the spatial moment dynamics approximation can depend on the choice of closure, as well as the region of parameter space under consideration~\cite{murrell2004moment,raghib2011multiscale,binny2016collective}. In some situations, the configuration of triples of agents may impact population-level behaviour in ways that are not captured by the pair density. In such cases, a more accurate approximation may be obtained by closing at the level of the third moment~\cite{kaito2015beyond}.

A derivation of the dynamics of the $k^\mathrm{th}$ moment for a general, multi-type ABM was given by Plank et al.~\cite{plank2015spatial}. This includes situations where the rates of motility, proliferation and death, and transitions between types, are enhanced or inhibited by homotypic or heterotypic neighbours via arbitrary interaction, movement and dispersal kernels. The derivation relies on an assumption that the effects of neighbours are additive and does not apply to situations where rates are nonlinear functions of the contributions from neighbours, e.g.~\cite{finkelshtein2013establishment}. 


\subsection{Applications of non-mean-field models} \label{sec:spatial_moments_applications}

{\highlight  The framework accounting for correlations between neighbouring lattice sites outlined in Section \ref{sec:spatial_structure_lattice_based} has also been applied on more general network topologies~\cite{keeling2005networks}. In both a regular lattice-based model and a network model, each node is in one of a set of possible states at any given time. However, in a lattice-based model in $d$ dimensions, the set of nodes adjacent to node $k$ simply consists of the $2d$ nearest lattice sites along each of the $d$ orthogonal coordinate directions (sometimes called the {\em von Neumann neighbourhood}). In a network model, the neighbourhood of node $k$ is the set of nodes that are connected to node $k$ by an edge. In general, the neighbourhood will depend on the network architecture and can be of arbitrary size. 

In the exclusion process model described in Section \ref{sec:spatial_structure_lattice_based}, the state space for each node is either empty (represented by $U_{i,j} = 0$) or occupied (represented by $U_{i,j} = 1$). An empty node may transition to the occupied state at a rate that depends on the number of neighbouring nodes that are occupied (either via a movement or a proliferation event). Other processes can be modelled in an analogous way using different state spaces. For example, models of infectious-disease dynamics canonically categorise individuals as either susceptible ($S$), infectious ($I$) or recovered ($R$)~\cite{kermack1927contribution}. A susceptible node may transition to the infectious state at a rate that depends on the number of infectious neighbours. Thus, the rate of new infections is determined by the number of pairs $[SI]$ of adjacent nodes where one node is susceptible and the other is infectious. 

The mean-field assumption is that $[SI]$ is equal to the product of the number of susceptible nodes $[S]$ and the number of infectious nodes $[I]$, which leads to a classical SIR model~\cite{anderson1991infectious}. An alternative approach is to express the dynamics of $[SI]$ in terms of the number of connected triples $[SSI]$ and $[ISI]$, and then use a moment closure to approximate these triples in terms of pairs and singles~\cite{keeling1997correlation,keeling1999effects}. This approach retains some information about the network architecture via correlations in the disease status of adjacent individuals. There is an extensive literature on moment closure approximations for network epidemic models (e.g.~\cite{kiss2017mathematics,sharkey2011deterministic,taylor2012markovian,pellis2015exact,leng2020improving}) but we do not describe this further here.}

Early lattice-free applications of spatial moment dynamics were mainly in plant ecology, where the spatial logistic model has been used to describe the effects of local competition and dispersal of seedlings~\cite{bolker1997using,bolker1999spatial,law2000dynamical}. The second spatial moment has the appealing property that it captures the local agent density from an individual ``plant's-eye view'', which can be quite different from the large-scale average density when there is significant spatial structure~\cite{purves2002fine,llambi2004temporal}. In a clustered population, the average individual experiences a higher local density than the large-scale average density, and vice versa in a population with regular structure.  

The basic lattice-free spatial moments framework has been extended to include density-dependent proliferation~\cite{lewis2000spread}, multi-species plant communities~\cite{dieckmann2000relaxation,murrell2003heteromyopia}, and size-structured populations~\cite{murrell2009emergent,adams2013growth}. The framework can also be extended to situations where agents can switch types~\cite{plank2015spatial}, which has been used to model infectious diseases whspatial moments frameworkere agents may transition from susceptible to infectious and subsequently recovered states~\cite{bolker1999analytic,brown2004effects}. It has also been used to model mutations, by allowing daughter agents to be a different type to their parent~\cite{champagnat2006unifying,champagnat2007invasion}. 

Because it relies an assumption that probabilities of events depend continuously on distances between agent centres and are additive across neighbours, the lattice-free spatial moments framework does not lend itself to modelling strict volume exclusion or hard sphere type interactions. Nevertheless, by adding motility, the approach has also been applied to animal ecology, including environmental interactions~\cite{murrell2000beetles}, predator-prey interactions~\cite{murrell2005local,barraquand2012evolutionarily,barraquand2013scaling,surendran2020small}, chase-escape interactions~\cite{surendran2019spatial}, and herding behaviour~\cite{binny2020living}. Binny et al.~\cite{binny2015spatial} incorporated a neighbour-dependent directional bias mechanism and used this to model crowding effects in experimental cell populations~\cite{binny2016spatial}. Browning et al.~\cite{browning2018inferring,browning2020identifying} used data from experimental cell populations to estimate the strength of neighbour-dependent effects on motility and proliferation. 

The majority of work on spatial point process models has focused on the spatially stationary case. Some studies have extended the framework to the {\em non-stationary} case~\cite{lewis2000modeling,omelyan2019spatially,plank2020small}, where the mean agent density varies with location. This is necessary for considering applications such as biological invasions~\cite{shigesada1997biological}, species range shifts~\cite{hurford2019skewed}, wound healing~\cite{maini2004traveling} and embryogenesis~\cite{binder2012spatial}. However, outside special cases and simulation-based studies, theory for the effects of local spatial structure in non-stationary populations is relatively underdeveloped. There are other approaches to accounting for spatial correlations that avoid the need for a closure approximation. These include working directly with the stochastic process~\cite{blath2007coexistence} and using a perturbation expansion in a small parameter $\epsilon$ representing, for example, the inverse of the spatial scale of interactions~\cite{ovaskainen2006asymptotically,ovaskainen2014general} or the excluded volume of hard spheres~\cite{bruna2012excluded,bruna2012diffusion} (see Section~\ref{sec:crowding_lattice_free}).


\section{Discussion and future challenges}

In this review we have described a range of mathematical and computational models for the spatiotemporal dynamics of populations of individual agents, with different mechanisms governing individual-level processes, including neighbour-independent and neighbour-dependent motility, proliferation and death. A particular emphasis has been to explore how continuum-limit equations approximating the macroscopic, population-level behaviour can be derived from different classes of ABM.

We have seen how including agent-agent interactions in an ABM can give rise to different terms in the continuum-limit equation. For example, including volume exclusion in a lattice-based ABM does not affect the linear diffusion term for random motility, but introduces a nonlinearity into the advection term for directed motility (Section \ref{sec:crowding_lattice_based}). Including other interactions, such as agent-agent adhesion or repulsion, can lead to a nonlinear diffusion term (Section \ref{sec:adhesion_repulsion}). Proliferation in a lattice-based ABM with volume exclusion can be described by a logistic growth term in the mean-field limit (Section \ref{sec:crowding_lattice_based_with_proliferation}). In contrast, in a lattice-free model, volume exclusion introduces a nonlinearity into the diffusion term (Section \ref{sec:crowding_lattice_free}) and proliferation is generally slower than under the logistic model when density is high due to the irregular spacing of agents (Section \ref{sec:crowding_lattice_free_with_proliferation}). Going beyond the mean-field by accounting for pairwise spatial correlations in agent locations can increase or decrease the net population growth rate relative to the mean-field model, depending on the nature and spatial scale of the interactions (Section \ref{sec:spatial_structure}). These insights are valuable because they reveal general relationships between individual-level mechanisms and population-level behaviour and highlight situations where population-level data may be insufficient to distinguish between alternative mechanistic hypotheses.  

Designing mechanistic mathematical models generally involves trade-offs between including sufficient complexity to answer the research questions of interest, and not over-complicating the model with irrelevant detail that obscures key mechanisms. Having a systematic framework for approximating the dynamics of stochastic ABMs with dynamical equations helps modellers and practitioners navigate these trade-offs. For example, it helps understand how, and under what circumstances, a particular individual-level mechanism will significantly impact population-level, observable behaviour, which assumptions and parameters are likely to sensitively affect model outputs, and when and why model approximations may break down.  

While this review attempts to cover key concepts relating to interacting random walk models and developments in this field, there are many ongoing challenges. For example, questions relating to fitting interacting random walk models with empirical data has become an important focus of research in recent years. This includes the development and application of methods for parameter identifiability and fitting and uncertainty quantification. Tools from both Bayesian (e.g.~Markov chain Monte Carlo and approximate Bayesian computation)~\cite{sisson2007sequential,toni2009approximate,sunnaaker2013approximate} and frequentist (e.g.~maximum likelihood and profile likelihood)~\cite{Kreutz2013,simpson2021profile,shuttleworth2024empirical} statistical inference may be applicable. Understanding how to implement these tools efficiently is important because stochastic ABMs typically involve a higher computational cost than deterministic continuum-limit models.  Practical parameter identifiability and experimental design can also support establishment of data collection protocols to ensure that appropriate summary statistics are collected to inform robust parameter estimates. 

Another active research area relates to model design and model selection. The development and implementation of suitable statistical tools in conjunction with population-level empirical data to robustly distinguish between putative individual-level mechanisms is an active avenue of research. This has the potential to enhance understanding of biological mechanisms, or understand when available data is insufficient to distinguish alternative biological hypotheses. Practical questions, such as whether to use a lattice-based or lattice-free model, or what kinds of individual-level mechanisms should be explicitly incorporated, are ongoing challenges in the field.  As we have illustrated in this work, it is possible to work with different stochastic models that have the same continuum-limit mean-field description, indicating that care is needed when identifying parameters and mechanisms from population-level data. Combining population-level and individual-level data may help distinguish between different individual-level mechanisms~\cite{Carr2021}.

Computational challenges, especially related to parameter identifiability and parameter estimation, have motivated the development of surrogate models that enable computationally efficient simulation ~\cite{SIMPSON2022Reliable}. This facilitates parameter estimation for complicated models that have high-dimensional parameter spaces. A related area of current research is the development of accurate moment closure methods that go beyond the mean-field and enable pairwise interactions to be included in models. Further, the development of equation-learning methods~\cite{Brunton2016} that enable the learning of accurate and interpretable ABMs and their coarse-grained approximations directly from data~\cite{Nardini2021,Messenger2022b} is an exciting area for future development. Constructing such models from noisy data, including single-agent tracking data and phenotypic heterogeneity is a key challenge~\cite{Messenger2022a}. 

Another challenge in the field is for modellers working in different application areas to learn from parallel developments in their respective areas. In some ecological contexts, the relative infrequency of interactions owing to relatively low population densities means that quantifying and understanding interactions can be difficult. In contrast, high-density environments are common in \textit{in vitro} cell biology, which means that methodologies in cell biology applications may be able to inform parallel developments in ecology. Similarly, applications in ecology may often involve modelling environmental heterogeneities (e.g. different land use patterns). These features are typically not encountered in \textit{in vitro} cell biology applications, yet they are clearly important for \textit{in vivo} applications.  Therefore, these two seemingly disparate application areas have much to learn from each other.


\subsection*{Data availability}

Code to reproduce the results in this paper is publicly available in the repository at \url{https://github.com/michaelplanknz/interacting-random-walk-models} and permanently archived at \url{https://doi.org/10.5281/zenodo.14037905}. 


\subsection*{Acknowledgements}  
MJP is supported by Te P\=unaha Matatini, Centre of Research Excellence in Complex Systems. MJS is supported by the Australian Research Council (DP230100025, CE230100001). REB is supported by a grant from the Simons Foundation (MP-SIP-00001828). We are grateful to the mathematical research institute MATRIX in Australia where part of this work was performed. We thank three anonymous reviewers for helpful comments.



\begin{thebibliography}{100}

\bibitem{kot1996dispersal}
Kot M, Lewis MA, van~den Driessche P.
\newblock Dispersal data and the spread of invading organisms.
\newblock Ecology. 1996;77(7):2027-42.

\bibitem{bolker1997using}
Bolker B, Pacala SW.
\newblock Using moment equations to understand stochastically driven spatial pattern formation in ecological systems.
\newblock Theoretical Population Biology. 1997;52(3):179-97.

\bibitem{simpson2007simulating}
Simpson MJ, Merrifield A, Landman KA, Hughes BD.
\newblock Simulating invasion with cellular automata: Connecting cell-scale and population-scale properties.
\newblock Physical Review E. 2007;76(2):021918.

\bibitem{Schmeiser2009}
Anguige K, Schmeiser C.
\newblock A one-dimensional model of cell diffusion and aggregation, incorporating volume filling and cell-to-cell adhesion.
\newblock Journal of Mathematical Biology. 2009;58:395â€“427.

\bibitem{codling2007group}
Codling EA, Pitchford JW, Simpson SD.
\newblock Group navigation and the ``many-wrongs principle'' in models of animal movement.
\newblock Ecology. 2007;88(7):1864-70.

\bibitem{murrell2005local}
Murrell DJ.
\newblock Local spatial structure and predator-prey dynamics: counterintuitive effects of prey enrichment.
\newblock American Naturalist. 2005;166:354-67.

\bibitem{berg1983random}
Berg HC.
\newblock Random walks in biology.
\newblock Princeton University Press; 1983.

\bibitem{okubo1989spatial}
Okubo A, Maini PK, Williamson MH, Murray JD.
\newblock On the spatial spread of the grey squirrel in Britain.
\newblock Proceedings of the Royal Society B. 1989;238(1291):113-25.

\bibitem{codling2008random}
Codling EA, Plank MJ, Benhamou S.
\newblock Random walk models in biology.
\newblock Journal of the Royal Society Interface. 2008;5(25):813-34.

\bibitem{weaver1948science}
Weaver W.
\newblock Science and complexity.
\newblock American Scientist. 1948;36(4):536-44.

\bibitem{pearson2005problem}
Pearson K.
\newblock The problem of the random walk.
\newblock Nature. 1905;72:294.

\bibitem{rayleigh2005problem}
Rayleigh.
\newblock The problem of the random walk.
\newblock Nature. 1905;72:318.

\bibitem{casquilho2015introduction}
Casquilho JP, Teixcira PIC.
\newblock Introduction to Statistical Physics and to Computer Simulations.
\newblock Cambridge University Press; 2015.

\bibitem{aubert2006cellular}
Aubert M, Badoual M, Fereol S, Christov C, Grammaticos B.
\newblock A cellular automaton model for the migration of glioma cells.
\newblock Physical Biology. 2006;3(2):93.

\bibitem{fernando2010nonlinear}
Fernando AE, Landman KA, Simpson MJ.
\newblock Nonlinear diffusion and exclusion processes with contact interactions.
\newblock Physical Review E. 2010;81(1):011903.

\bibitem{Engblom:2009:SSR}
Engblom S, Ferm L, Hellander A, Lotstedt P.
\newblock Simulation of stochastic reaction-diffusion processes on unstructured meshes.
\newblock SIAM Journal on Scientific Computing. 2009;31(3):1774-97.

\bibitem{Lotstedt:2015:SOS}
L\"{o}tstedt P, Meinecke L.
\newblock Simulation of stochastic diffusion via first exit times.
\newblock Journal of Computational Physics. 2015;300:862-86.

\bibitem{lin1974mathematics}
Lin CC, Segel LA.
\newblock Mathematics applied to deterministic problems in the natural sciences.
\newblock Macmillan; 1974.

\bibitem{Morton2005}
Morton KW, Mayers DF.
\newblock Numerical solution of partial differential equations.
\newblock Cambridge University Press; 2005.

\bibitem{othmer1997aggregation}
Othmer HG, Stevens A.
\newblock Aggregation, blowup, and collapse: the ABC's of taxis in reinforced random walks.
\newblock SIAM Journal on Applied Mathematics. 1997;57(4):1044-81.

\bibitem{murray2003mathematical}
Murray JD.
\newblock Mathematical biology I: an introduction.
\newblock Springer; 2003.

\bibitem{okubo2001diffusion}
Okubo A, Levin SA.
\newblock Diffusion and ecological problems: modern perspectives.
\newblock Springer; 2001.

\bibitem{mardia2009directional}
Mardia KV, Jupp PE.
\newblock Directional statistics.
\newblock John Wiley \& Sons; 2009.

\bibitem{cheung2007animal}
Cheung A, Zhang S, Stricker C, Srinivasan MV.
\newblock Animal navigation: the difficulty of moving in a straight line.
\newblock Biological Cybernetics. 2007;97:47-61.

\bibitem{codling2010diffusion}
Codling EA, Bearon RN, Thorn GJ.
\newblock Diffusion about the mean drift location in a biased random walk.
\newblock Ecology. 2010;91(10):3106-13.

\bibitem{montroll1984wonderful}
Montroll EW, Shlesinger MF.
\newblock On the wonderful world of random walks.
\newblock In: Lebowitz JL, Montrol EW, editors. Nonequilibrium phenomena. II: From stochastics to hydrodynamics. North--Holland; 1984. p. 1-121.

\bibitem{grimmett2001probability}
Grimmett G, Stirzaker D.
\newblock Probability and random processes.
\newblock Oxford University Press; 2001.

\bibitem{weeks1996anomalous}
Weeks ER, Urbach JS, Swinney HL.
\newblock Anomalous diffusion in asymmetric random walks with a quasi-geostrophic flow example.
\newblock Physica D: Nonlinear Phenomena. 1996;97(1-3):291-310.

\bibitem{edwards2007revisiting}
Edwards AM, Phillips RA, Watkins NW, Freeman MP, Murphy EJ, Afanasyev V, et~al.
\newblock Revisiting L{\'e}vy flight search patterns of wandering albatrosses, bumblebees and deer.
\newblock Nature. 2007;449(7165):1044-8.

\bibitem{viswanathan2011physics}
Viswanathan GM, Da~Luz MGE, Raposo EP, Stanley HE.
\newblock The physics of foraging: an introduction to random searches and biological encounters.
\newblock Cambridge University Press; 2011.

\bibitem{kareiva1983analyzing}
Kareiva PM, Shigesada N.
\newblock Analyzing insect movement as a correlated random walk.
\newblock Oecologia. 1983;56:234-8.

\bibitem{othmer1988models}
Othmer HG, Dunbar SR, Alt W.
\newblock Models of dispersal in biological systems.
\newblock Journal of Mathematical Biology. 1988;26(3):263-98.

\bibitem{othmer2000diffusion}
Othmer HG, Hillen T.
\newblock The diffusion limit of transport equations derived from velocity-jump processes.
\newblock SIAM Journal on Applied Mathematics. 2000;61(3):751-75.

\bibitem{bellomo2017active}
Bellomo N, Degond P, Tadmor E.
\newblock Active Particles, Volume 1: Advances in Theory, Models, and Applications.
\newblock Birkh{\"a}user; 2017.

\bibitem{patlak1953random}
Patlak CS.
\newblock Random walk with persistence and external bias.
\newblock Bulletin of Mathematical Biophysics. 1953;15:311-38.

\bibitem{hill1997biased}
Hill NA, H{\"a}der DP.
\newblock A biased random walk model for the trajectories of swimming micro-organisms.
\newblock Journal of Theoretical Biology. 1997;186(4):503-26.

\bibitem{bearon2000modelling}
Bearon RN, Pedley TJ.
\newblock Modelling run-and-tumble chemotaxis in a shear flow.
\newblock Bulletin of Mathematical Biology. 2000;62(4):775-91.

\bibitem{othmer2002diffusion}
Othmer HG, Hillen T.
\newblock The diffusion limit of transport equations II: chemotaxis equations.
\newblock SIAM Journal on Applied Mathematics. 2002;62(4):1222-50.

\bibitem{stokes1991analysis}
Stokes CL, Lauffenburger DA.
\newblock Analysis of the roles of microvessel endothelial cell random motility and chemotaxis in angiogenesis.
\newblock Journal of Theoretical Biology. 1991;152(3):377-403.

\bibitem{plank2004lattice}
Plank MJ, Sleeman BD.
\newblock Lattice and non-lattice models of tumour angiogenesis.
\newblock Bulletin of Mathematical Biology. 2004;66:1785-819.

\bibitem{bovet1988spatial}
Bovet P, Benhamou S.
\newblock Spatial analysis of animals' movements using a correlated random walk model.
\newblock Journal of Theoretical Biology. 1988;131(4):419-33.

\bibitem{codling2004random}
Codling EA, Hill NA, Pitchford JW, Simpson SD.
\newblock Random walk models for the movement and recruitment of reef fish larvae.
\newblock Marine Ecology Progress Series. 2004;279:215-24.

\bibitem{fa2013generalized}
Fa KS, Wang KG.
\newblock Generalized Klein--Kramers equation: solution and application.
\newblock Journal of Statistical Mechanics: Theory and Experiment. 2013;2013(09):P09021.

\bibitem{goldstein1951diffusion}
Goldstein S.
\newblock On diffusion by discontinuous movements, and on the telegraph equation.
\newblock Quarterly Journal of Mechanics and Applied Mathematics. 1951;4(2):129-56.

\bibitem{kac1974stochastic}
Kac M.
\newblock A stochastic model related to the telegrapher's equation.
\newblock Rocky Mountain Journal of Mathematics. 1974;4(3):497-509.

\bibitem{erban2004individual}
Erban R, Othmer HG.
\newblock From individual to collective behavior in bacterial chemotaxis.
\newblock SIAM Journal on Applied Mathematics. 2004;65(2):361-91.

\bibitem{erban2005signal}
Erban R, Othmer HG.
\newblock From signal transduction to spatial pattern formation in E. coli: a paradigm for multiscale modeling in biology.
\newblock Multiscale Modeling and Simulation. 2005;3(2):362-94.

\bibitem{doi1976stochastic}
Doi M.
\newblock Stochastic theory of diffusion-controlled reaction.
\newblock Journal of Physics A: Mathematical and General. 1976;9(9):1479.

\bibitem{peliti1985path}
Peliti L.
\newblock Path integral approach to birth-death processes on a lattice.
\newblock Journal de Physique. 1985;46(9):1469-83.

\bibitem{isaacson2008relationship}
Isaacson SA.
\newblock Relationship between the reaction--diffusion master equation and particle tracking models.
\newblock Journal of Physics A: Mathematical and Theoretical. 2008;41(6):065003.

\bibitem{skellam1951dispersal}
Skellam JG.
\newblock Random dispersal in theoretical populations.
\newblock Biometrika. 1951;38:196-218.

\bibitem{durrett1994stochastic}
Durrett R, Levin SA.
\newblock Stochastic spatial models: a user's guide to ecological applications.
\newblock Philosophical Transactions of the Royal Society of London Series B: Biological Sciences. 1994;343(1305):329-50.

\bibitem{liggett1999stochastic}
Liggett TM.
\newblock Stochastic interacting systems: contact, voter and exclusion processes.
\newblock Berlin: Springer; 1999.

\bibitem{Simpson2009}
Simpson MJ, Landman KA, Hughes BD.
\newblock Multi-species simple exclusion processes.
\newblock Physica A: Statistical Mechanics and its Applications. 2009;388(4):399-406.

\bibitem{wei2000single}
Wei QH, Bechinger C, Leiderer P.
\newblock Single-file diffusion of colloids in one-dimensional channels.
\newblock Science. 2000;287(5453):625-7.

\bibitem{schonherr2004exclusion}
Sch{\"o}nherr G, Sch{\"u}tz GM.
\newblock Exclusion process for particles of arbitrary extension: hydrodynamic limit and algebraic properties.
\newblock Journal of Physics A: Mathematical and General. 2004;37(34):8215.

\bibitem{lakatos2006hydrodynamic}
Lakatos G, O'Brien J, Chou T.
\newblock Hydrodynamic mean-field solutions of 1D exclusion processes with spatially varying hopping rates.
\newblock Journal of Physics A: Mathematical and General. 2006;39(10):2253.

\bibitem{grabsch2024tracer}
Grabsch A, B{\'e}nichou O.
\newblock Tracer diffusion beyond Gaussian behavior: explicit results for general single-file systems.
\newblock Physical Review Letters. 2024;132(21):217101.

\bibitem{SimpsonPathlines}
Simpson MJ, Landman KA, Hughes BD.
\newblock Pathlines in exclusion processes.
\newblock Physical Review E. 2009;79:031920.

\bibitem{Simpson2010}
Simpson MJ, Landman KA, Hughes BD.
\newblock Cell invasion with proliferation mechanisms motivated by time-lapse data.
\newblock Physica A: Statistical Mechanics and its Applications. 2010;389(18):3779-90.

\bibitem{baker2010correcting}
Baker RE, Simpson MJ.
\newblock Correcting mean-field approximations for birth-death-movement processes.
\newblock Physical Review E. 2010;82(4):041905.

\bibitem{khain2011collective}
Khain E, Katakowski M, Hopkins S, Szalad A, Zheng X, Jiang F, et~al.
\newblock Collective behavior of brain tumor cells: the role of hypoxia.
\newblock Physical Review E. 2011;83(3):031920.

\bibitem{coscoy2007statistical}
Coscoy S, Huguet E, Amblard F.
\newblock Statistical analysis of sets of random walks: how to resolve their generating mechanism.
\newblock Bulletin of Mathematical Biology. 2007;69(8):2467-92.

\bibitem{ziff2009capture}
Ziff RM, Majumdar SN, Comtet A.
\newblock Capture of particles undergoing discrete random walks.
\newblock Journal of Chemical Physics. 2009;130(20).

\bibitem{dyson2012macroscopic}
Dyson L, Maini PK, Baker RE.
\newblock Macroscopic limits of individual-based models for motile cell populations with volume exclusion.
\newblock Physical Review E. 2012;86(3):031903.

\bibitem{dyson2015importance}
Dyson L, Baker RE.
\newblock The importance of volume exclusion in modelling cellular migration.
\newblock Journal of Mathematical Biology. 2015;71:691-711.

\bibitem{bruna2012excluded}
Bruna M, Chapman SJ.
\newblock Excluded-volume effects in the diffusion of hard spheres.
\newblock Physical Review E. 2012;85(1):011103.

\bibitem{bruna2012diffusion}
Bruna M, Chapman SJ.
\newblock Diffusion of multiple species with excluded-volume effects.
\newblock Journal of Chemical Physics. 2012;137(20):204116.

\bibitem{bruna2014diffusion}
Bruna M, Chapman SJ.
\newblock Diffusion of finite-size particles in confined geometries.
\newblock Bulletin of Mathematical Biology. 2014;76:947-82.

\bibitem{Treloar2011}
Treloar KK, Simpson MJ, McCue SW.
\newblock Velocity-jump models with crowding effects.
\newblock Physical Review E. 2011;84:061920.

\bibitem{alert2020physical}
Alert R, Trepat X.
\newblock Physical models of collective cell migration.
\newblock Annual Review of Condensed Matter Physics. 2020;11(1):77-101.

\bibitem{plank2012models}
Plank MJ, Simpson MJ.
\newblock Models of collective cell behaviour with crowding effects: comparing lattice-based and lattice-free approaches.
\newblock Journal of the Royal Society Interface. 2012;9(76):2983-96.

\bibitem{tremel2009cell}
Tremel A, Cai A, Tirtaatmadja N, Hughes BD, Stevens GW, Landman KA, et~al.
\newblock Cell migration and proliferation during monolayer formation and wound healing.
\newblock Chemical Engineering Science. 2009;64(2):247-53.

\bibitem{plank2013lattice}
Plank MJ, Simpson MJ.
\newblock Lattice-free models of cell invasion: discrete simulations and travelling waves.
\newblock Bulletin of Mathematical Biology. 2013;75(11):2150-66.

\bibitem{irons2015lattice}
Irons C, Plank MJ, Simpson MJ.
\newblock Lattice-free models of directed cell motility.
\newblock Physica A: Statistical Mechanics and its Applications. 2015;442:110-21.

\bibitem{deroulers2009modeling}
Deroulers C, Aubert M, Badoual M, Grammaticos B.
\newblock Modeling tumor cell migration: From microscopic to macroscopic models.
\newblock Physical Review E. 2009;79(3):031917.

\bibitem{johnston2012mean}
Johnston ST, Simpson MJ, Baker RE.
\newblock Mean-field descriptions of collective migration with strong adhesion.
\newblock Physical Review E. 2012;85(5):051922.

\bibitem{Thompson2012}
Thompson RN, Yates CA, Baker RE.
\newblock Modelling cell migration and adhesion during development.
\newblock Bulletin of Mathematical Biology. 2012;74(12):2793-809.

\bibitem{SimpsonMcElwainUpton2010}
Simpson MJ, Towne C, McElwain DLS, Upton Z.
\newblock Migration of breast cancer cells: Understanding the roles of volume exclusion and cell-to-cell adhesion.
\newblock Physical Review E. 2010;82:041901.

\bibitem{Johnston2013}
Johnston ST, Simpson MJ, Plank MJ.
\newblock Lattice-free descriptions of collective motion with crowding and adhesion.
\newblock Physical Review E. 2013;88:062720.

\bibitem{Simpsonmeso}
Simpson MJ, Landman KA, Hughes BD, Fernando AE.
\newblock A model for mesoscale patterns in motile populations.
\newblock Physica A: Statistical Mechanics and its Applications. 2010;389(7):1412-24.

\bibitem{GAVAGNIN201991}
Gavagnin E, Ford MJ, Mort RL, Rogers T, Yates CA.
\newblock The invasion speed of cell migration models with realistic cell cycle time distributions.
\newblock Journal of Theoretical Biology. 2019;481:91-9.

\bibitem{SIMPSON2018}
Simpson MJ, Jin W, Vittadello ST, Tambyah TA, Ryan JM, Gunasingh G, et~al.
\newblock Stochastic models of cell invasion with fluorescent cell cycle indicators.
\newblock Physica A: Statistical Mechanics and its Applications. 2018;510:375-86.

\bibitem{reina2012lattice}
Reina-Romo E, G{\'o}mez-Benito MJ, Dom{\'\i}nguez J, Garc{\'\i}a-Aznar JM.
\newblock A lattice-based approach to model distraction osteogenesis.
\newblock Journal of Biomechanics. 2012;45(16):2736-42.

\bibitem{adams2013growth}
Adams T, Holland EP, Law R, Plank MJ, Raghib M.
\newblock On the growth of locally interacting plants: differential equations for the dynamics of spatial moments.
\newblock Ecology. 2013;94:2732-43.

\bibitem{molina2015analyzing}
Molina MM, Moreno-Armend{\'a}riz MA, Mora JCST.
\newblock Analyzing the spatial dynamics of a prey--predator lattice model with social behavior.
\newblock Ecological Complexity. 2015;22:192-202.

\bibitem{durrett1998spatial}
Durrett R, Levin S.
\newblock Spatial aspects of interspecific competition.
\newblock Theoretical Population Biology. 1998;53(1):30-43.

\bibitem{ying2014species}
Ying Z, Liao J, Wang S, Lu H, Liu Y, Ma L, et~al.
\newblock Species coexistence in a lattice-structured habitat: Effects of species dispersal and interactions.
\newblock Journal of Theoretical Biology. 2014;359:184-91.

\bibitem{windus2007allee}
Windus A, Jensen HJ.
\newblock Allee effects and extinction in a lattice model.
\newblock Theoretical Population Biology. 2007;72(4):459-67.

\bibitem{rhodes1997epidemic}
Rhodes CJ, Anderson RM.
\newblock Epidemic thresholds and vaccination in a lattice model of disease spread.
\newblock Theoretical Population Biology. 1997;52(2):101-18.

\bibitem{keeling1999effects}
Keeling MJ.
\newblock The effects of local spatial structure on epidemiological invasions.
\newblock Proceedings of the Royal Society B. 1999;266(1421):859-67.

\bibitem{jin2021mathematical}
Jin W, Spoerri L, Haass NK, Simpson MJ.
\newblock Mathematical model of tumour spheroid experiments with real-time cell cycle imaging.
\newblock Bulletin of Mathematical Biology. 2021;83:1-23.

\bibitem{Baker2010}
Baker RE, Yates CA, Erban R.
\newblock From microscopic to macroscopic descriptions of cell migration on growing domains.
\newblock Bulletin of Mathematical Biology. 2010;72:719-62.

\bibitem{Yates2012}
Yates CA, Baker RE, Erban R, Maini PK.
\newblock Going from microscopic to macroscopic on nonuniform growing domains.
\newblock Physical Review E. 2012;86:021921.

\bibitem{Yates2013}
Yates CA, Baker RE.
\newblock Importance of the Voronoi domain partition for position-jump reaction-diffusion processes on nonuniform rectilinear lattices.
\newblock Physical Review E. 2013;88:054701.

\bibitem{Binder2008}
Binder BJ, Landman KA, Simpson MJ, Mariani M, Newgreen DF.
\newblock Modeling proliferative tissue growth: A general approach and an avian case study.
\newblock Physical Review E. 2008;78:031912.

\bibitem{YATES2014}
Yates CA.
\newblock Discrete and continuous models for tissue growth and shrinkage.
\newblock Journal of Theoretical Biology. 2014;350:37-48.

\bibitem{Simpson2015}
Simpson MJ, Baker RE.
\newblock Exact calculations of survival probability for diffusion on growing lines, disks, and spheres: the role of dimension.
\newblock Journal of Chemical Physics. 2015;143(9):094109.

\bibitem{kirkwood1935statistical}
Kirkwood JG.
\newblock Statistical mechanics of fluid mixtures.
\newblock Journal of Chemical Physics. 1935;3:300-13.

\bibitem{matsuda1992statistical}
Matsuda H, Ogita N, Sasaki A, Sato K.
\newblock Statistical mechanics of population: the lattice {L}otka-{V}olterra model.
\newblock Progress of Theoretical Physics. 1992;88:1035-49.

\bibitem{keeling2000multiplicative}
Keeling MJ.
\newblock Multiplicative moments and measures of persistence in ecology.
\newblock Journal of Theoretical Biology. 2000;205(2):269-81.

\bibitem{lewis2000modeling}
Lewis MA, Pacala S.
\newblock Modeling and analysis of stochastic invasion processes.
\newblock Journal of Mathematical Biology. 2000;41(5):387-429.

\bibitem{dieckmann2000relaxation}
Dieckmann U, Law R.
\newblock Relaxation projections and the method of moments.
\newblock In: Dieckmann U, Law R, Metz JAJ, editors. The Geometry of ecological interactions: simplifying spatial complexity. Cambridge, UK: Cambridge University Press; 2000. p. 412-55.

\bibitem{bolker2003combining}
Bolker BM.
\newblock Combining endogenous and exogenous spatial variability in analytical population models.
\newblock Theoretical Population Biology. 2003;64(3):255-70.

\bibitem{murrell2004moment}
Murrell DJ, Dieckmann U, Law R.
\newblock On moment closures for population dynamics in continuous space.
\newblock Journal of Theoretical Biology. 2004;229(3):421-32.

\bibitem{raghib2011multiscale}
Raghib M, Hill NA, Dieckmann U.
\newblock A multiscale maximum entropy moment closure for locally regulated space--time point process models of population dynamics.
\newblock Journal of Mathematical Biology. 2011;62(5):605-53.

\bibitem{simpson2011corrected}
Simpson MJ, Baker RE.
\newblock Corrected mean-field models for spatially dependent advection-diffusion-reaction phenomena.
\newblock Physical Review E. 2011;83(5):051922.

\bibitem{Markham2013}
Markham DC, Simpson MJ, Maini PK, Gaffney EA, Baker RE.
\newblock Incorporating spatial correlations into multispecies mean-field models.
\newblock Physical Review E. 2013;88:052713.

\bibitem{JOHNSTON201581}
Johnston ST, Simpson MJ, Baker RE.
\newblock Modelling the movement of interacting cell populations: a moment dynamics approach.
\newblock Journal of Theoretical Biology. 2015;370:81-92.

\bibitem{ellner2001pair}
Ellner SP.
\newblock Pair approximation for lattice models with multiple interaction scales.
\newblock Journal of Theoretical Biology. 2001;210:435-47.

\bibitem{ROSS2015}
Ross RJH, Yates CA, Baker RE.
\newblock Inference of cellâ€“cell interactions from population density characteristics and cell trajectories on static and growing domains.
\newblock Mathematical Biosciences. 2015;264:108-18.

\bibitem{Ross2016}
Ross RJH, Baker RE, Yates CA.
\newblock How domain growth is implemented determines the long-term behavior of a cell population through its effect on spatial correlations.
\newblock Physical Review E. 2016;94:012408.

\bibitem{iwasa2000}
Iwasa Y.
\newblock Lattice models and pair approximation in ecology.
\newblock In: Dieckmann U, Law R, Metz JAJ, editors. The geometry of ecological interactions: simplifying spatial complexity. Cambridge University Press; 2000. p. 227-47.

\bibitem{illian2008statistical}
Illian J, Penttinen A, Stoyan H, Stoyan D.
\newblock Statistical analysis and modelling of spatial point patterns.
\newblock Chichester: John Wiley \& Sons; 2008.

\bibitem{finkelshtein2009individual}
Finkelshtein D, Kondratiev Y, Kutoviy O.
\newblock Individual-based model with competition in spatial ecology.
\newblock SIAM Journal on Mathematical Analysis. 2009;41(1):297-317.

\bibitem{hastings2005spatial}
Hastings A, Cuddington K, Davies KF, Dugaw CJ, Elmendorf S, Freestone A, et~al.
\newblock The spatial spread of invasions: new developments in theory and evidence.
\newblock Ecology Letters. 2005;8(1):91-101.

\bibitem{law2003population}
Law R, Murrell DJ, Dieckmann U.
\newblock Population growth in space and time: spatial logistic equations.
\newblock Ecology. 2003;84(1):252-62.

\bibitem{bolker1999spatial}
Bolker BM, Pacala SW.
\newblock Spatial moment equations for plant competition: understanding spatial strategies and the advantages of short dispersal.
\newblock American Naturalist. 1999;153(6):575-602.

\bibitem{plank2015spatial}
Plank MJ, Law R.
\newblock Spatial point processes and moment dynamics in the life sciences: a parsimonious derivation and some extensions.
\newblock Bulletin of Mathematical Biology. 2015;77:586-613.

\bibitem{law2000dynamical}
Law R, Dieckmann U.
\newblock A dynamical system for neighborhoods in plant communities.
\newblock Ecology. 2000;81:2137-48.

\bibitem{binny2016collective}
Binny RN, James A, Plank MJ.
\newblock Collective cell behaviour with neighbour-dependent proliferation, death and directional bias.
\newblock Bulletin of Mathematical Biology. 2016;78:2277-301.

\bibitem{kaito2015beyond}
Kaito C, Dieckmann U, Sasaki A, Takasu F.
\newblock Beyond pairs: definition and interpretation of third-order structure in spatial point patterns.
\newblock Journal of Theoretical Biology. 2015;372:22-38.

\bibitem{finkelshtein2013establishment}
Finkelshtein D, Kondratiev Y, Kutoviy O.
\newblock Establishment and fecundity in spatial ecological models: statistical approach and kinetic equations.
\newblock Infinite Dimensional Analysis, Quantum Probability and Related Topics. 2013;16(02).

\bibitem{keeling2005networks}
Keeling MJ, Eames KTD.
\newblock Networks and epidemic models.
\newblock Journal of the Royal Society Interface. 2005;2(4):295-307.

\bibitem{kermack1927contribution}
Kermack WO, McKendrick AG.
\newblock A Contribution to the Mathematical Theory of Epidemics.
\newblock Proceedings of the Royal Society A. 1927;115(772):700-21.

\bibitem{anderson1991infectious}
Anderson RM, May RM.
\newblock Infectious diseases of humans: dynamics and control.
\newblock Oxford University Press; 1991.

\bibitem{keeling1997correlation}
Keeling MJ, Rand DA, Morris AJ.
\newblock Correlation models for childhood epidemics.
\newblock Proceedings of the Royal Society B. 1997;264(1385):1149-56.

\bibitem{kiss2017mathematics}
Kiss IZ, Miller JC, Simon PL.
\newblock Mathematics of epidemics on networks.
\newblock Springer; 2017.

\bibitem{sharkey2011deterministic}
Sharkey KJ.
\newblock Deterministic epidemic models on contact networks: Correlations and unbiological terms.
\newblock Theoretical Population Biology. 2011;79(4):115-29.

\bibitem{taylor2012markovian}
Taylor M, Simon PL, Green DM, House T, Kiss IZ.
\newblock From Markovian to pairwise epidemic models and the performance of moment closure approximations.
\newblock Journal of Mathematical Biology. 2012;64:1021-42.

\bibitem{pellis2015exact}
Pellis L, House T, Keeling MJ.
\newblock Exact and approximate moment closures for non-Markovian network epidemics.
\newblock Journal of Theoretical Biology. 2015;382:160-77.

\bibitem{leng2020improving}
Leng T, Keeling MJ.
\newblock Improving pairwise approximations for network models with susceptible-infected-susceptible dynamics.
\newblock Journal of Theoretical Biology. 2020;500:110328.

\bibitem{purves2002fine}
Purves DW, Law R.
\newblock Fine-scale spatial structure in a grassland community: quantifying the plant's-eye view.
\newblock Journal of Ecology. 2002;90:121-9.

\bibitem{llambi2004temporal}
Llambi LD, Law R, Hodge A.
\newblock Temporal changes in local spatial structure of late-successional species: establishment of an {A}ndean caulescent rosette plant.
\newblock Journal of Ecology. 2004;92:122-31.

\bibitem{lewis2000spread}
Lewis MA.
\newblock Spread rate for a nonlinear stochastic invasion.
\newblock Journal of Mathematical Biology. 2000;41(5):430-54.

\bibitem{murrell2003heteromyopia}
Murrell DJ, Law R.
\newblock Heteromyopia and the spatial coexistence of similar competitors.
\newblock Ecology Letters. 2003;6(1):48-59.

\bibitem{murrell2009emergent}
Murrell DJ.
\newblock On the emergent spatial structure of size-structured populations: when does self-thinning lead to a reduction in clustering?
\newblock Journal of Ecology. 2009;97:256-66.

\bibitem{bolker1999analytic}
Bolker BM.
\newblock Analytic models for the patchy spread of plant disease.
\newblock Bulletin of Mathematical Biology. 1999;61(5):849-74.

\bibitem{brown2004effects}
Brown DH, Bolker BM.
\newblock The effects of disease dispersal and host clustering on the epidemic threshold in plants.
\newblock Bulletin of Mathematical Biology. 2004;66(2):341-71.

\bibitem{champagnat2006unifying}
Champagnat N, Ferri{\`e}re R, M{\'e}l{\'e}ard S.
\newblock Unifying evolutionary dynamics: from individual stochastic processes to macroscopic models.
\newblock Theoretical Population Biology. 2006;69(3):297-321.

\bibitem{champagnat2007invasion}
Champagnat N, M{\'e}l{\'e}ard S.
\newblock Invasion and adaptive evolution for individual-based spatially structured populations.
\newblock Journal of Mathematical Biology. 2007;55(2):147-88.

\bibitem{murrell2000beetles}
Murrell DJ, Law R.
\newblock Beetles in fragmented woodlands: a formal framework for dynamics of movement in ecological landscapes.
\newblock Journal of Animal Ecology. 2000;69(3):471-83.

\bibitem{barraquand2012evolutionarily}
Barraquand F, Murrell DJ.
\newblock Evolutionarily stable consumer home range size in relation to resource demography and consumer spatial organization.
\newblock Theoretical Ecology. 2012;5(4):567-89.

\bibitem{barraquand2013scaling}
Barraquand F, Murrell DJ.
\newblock Scaling up predator--prey dynamics using spatial moment equations.
\newblock Methods in Ecology and Evolution. 2013;4:276-89.

\bibitem{surendran2020small}
Surendran A, Plank MJ, Simpson MJ.
\newblock Small-scale spatial structure affects predator-prey dynamics and coexistence.
\newblock Theoretical Ecology. 2020;13(4):537-50.

\bibitem{surendran2019spatial}
Surendran A, Plank MJ, Simpson MJ.
\newblock Spatial structure arising from chase-escape interactions with crowding.
\newblock Scientific Reports. 2019;9(1):14988.

\bibitem{binny2020living}
Binny RN, Law R, Plank MJ.
\newblock Living in groups: spatial-moment dynamics with neighbour-biased movements.
\newblock Ecological Modelling. 2020;415:108825.

\bibitem{binny2015spatial}
Binny RN, Plank MJ, James A.
\newblock Spatial moment dynamics for collective cell movement incorporating a neighbour-dependent directional bias.
\newblock Journal of the Royal Society Interface. 2015;12(106):20150228.

\bibitem{binny2016spatial}
Binny RN, Haridas P, James A, Law R, Simpson MJ, Plank MJ.
\newblock Spatial structure arising from neighbour-dependent bias in collective cell movement.
\newblock PeerJ. 2016;4:e1689.

\bibitem{browning2018inferring}
Browning AP, McCue SW, Binny RN, Plank MJ, Shah ET, Simpson MJ.
\newblock Inferring parameters for a lattice-free model of cell migration and proliferation using experimental data.
\newblock Journal of Theoretical Biology. 2018;437:251-60.

\bibitem{browning2020identifying}
Browning AP, Jin W, Plank MJ, Simpson MJ.
\newblock Identifying density-dependent interactions in collective cell behaviour.
\newblock Journal of The Royal Society Interface. 2020;17(165):20200143.

\bibitem{omelyan2019spatially}
Omelyan I, Kozitsky Y.
\newblock Spatially inhomogeneous population dynamics: beyond the mean field approximation.
\newblock Journal of Physics A: Mathematical and Theoretical. 2019;52(30):305601.

\bibitem{plank2020small}
Plank MJ, Simpson MJ, Binny RN.
\newblock Small-scale spatial structure influences large-scale invasion rates.
\newblock Theoretical Ecology. 2020;13(3):277-88.

\bibitem{shigesada1997biological}
Shigesada N, Kawasaki K.
\newblock Biological invasions: theory and practice.
\newblock Oxford: Oxford University Press; 1997.

\bibitem{hurford2019skewed}
Hurford A, Cobbold CA, Moln{\'a}r PK.
\newblock Skewed temperature dependence affects range and abundance in a warming world.
\newblock Proceedings of the Royal Society B. 2019;286(1908):20191157.

\bibitem{maini2004traveling}
Maini PK, McElwain DLS, Leavesley DI.
\newblock Traveling wave model to interpret a wound-healing cell migration assay for human peritoneal mesothelial cells.
\newblock Tissue Engineering. 2004;10(3-4):475-82.

\bibitem{binder2012spatial}
Binder BJ, Landman KA, Newgreen DF, Simkin JE, Takahashi Y, Zhang D.
\newblock Spatial analysis of multi-species exclusion processes: application to neural crest cell migration in the embryonic gut.
\newblock Bulletin of Mathematical Biology. 2012;74:474-90.

\bibitem{blath2007coexistence}
Blath J, Etheridge A, Meredith M.
\newblock Coexistence in locally regulated competing populations and survival of branching annihilating random walk.
\newblock Annals of Applied Probability. 2007:1474-507.

\bibitem{ovaskainen2006asymptotically}
Ovaskainen O, Cornell SJ.
\newblock Asymptotically exact analysis of stochastic metapopulation dynamics with explicit spatial structure.
\newblock Theoretical Population Biology. 2006;69(1):13-33.

\bibitem{ovaskainen2014general}
Ovaskainen O, Finkelshtein D, Kutoviy O, Cornell S, Bolker B, Kondratiev Y.
\newblock A general mathematical framework for the analysis of spatiotemporal point processes.
\newblock Theoretical Ecology. 2014;7:101-13.

\bibitem{sisson2007sequential}
Sisson SA, Fan Y, Tanaka MM.
\newblock Sequential Monte Carlo without likelihoods.
\newblock Proceedings of the National Academy of Sciences. 2007;104(6):1760-5.

\bibitem{toni2009approximate}
Toni T, Welch D, Strelkowa N, Ipsen A, Stumpf MP.
\newblock Approximate {B}ayesian computation scheme for parameter inference and model selection in dynamical systems.
\newblock Journal of the Royal Society Interface. 2009;6(31):187-202.

\bibitem{sunnaaker2013approximate}
Sunn{\aa}ker M, Busetto AG, Numminen E, Corander J, Foll M, Dessimoz C.
\newblock Approximate {B}ayesian computation.
\newblock PLoS Computational Biology. 2013;9(1):e1002803.

\bibitem{Kreutz2013}
Kreutz C, Raue A, Kaschek D, Timmer J.
\newblock Profile likelihood in systems biology.
\newblock FEBS Journal. 2013;280:2564-71.

\bibitem{simpson2021profile}
Simpson MJ, Browning AP, Drovandi C, Carr EJ, Maclaren OJ, Baker RE.
\newblock Profile likelihood analysis for a stochastic model of diffusion in heterogeneous media.
\newblock Proceedings of the Royal Society A. 2021;477(2250):20210214.

\bibitem{shuttleworth2024empirical}
Shuttleworth JG, Lei CL, Whittaker DG, Windley MJ, Hill AP, Preston SP, et~al.
\newblock Empirical quantification of predictive uncertainty due to model discrepancy by training with an ensemble of experimental designs: an application to ion channel kinetics.
\newblock Bulletin of Mathematical Biology. 2024;86(1):2.

\bibitem{Carr2021}
Carr MJ, Simpson MJ, Drovandi C.
\newblock Estimating parameters of a stochastic cell invasion model with fluorescent cell cycle labelling using approximate Bayesian computation.
\newblock Journal of the Royal Society Interface. 2021;18:20210362.

\bibitem{SIMPSON2022Reliable}
Simpson MJ, Baker RE, Buenzli PR, Nicholson R, Maclaren OJ.
\newblock Reliable and efficient parameter estimation using approximate continuum limit descriptions of stochastic models.
\newblock Journal of Theoretical Biology. 2022;549:111201.

\bibitem{Brunton2016}
Brunton SL, Proctor JL, Kutz JN.
\newblock Discovering governing equations from data by sparse identification of nonlinear dynamical systems.
\newblock Proceedings of the National Academy of Sciences. 2016;113:3932-7.

\bibitem{Nardini2021}
Nardini JT, Baker RE, Simpson MJ, Flores KB.
\newblock Learning differential equation models from stochastic agent-based model simulations.
\newblock Journal of the Royal Society Interface. 2021;18:20200987.

\bibitem{Messenger2022b}
Messenger DA, Bortz DM.
\newblock Learning mean-field equations from particle data using WSINDy.
\newblock Physica D: Nonlinear Phenomena. 2022;439:133406.

\bibitem{Messenger2022a}
Messenger DA, Wheeler GE, Liu X, Bortz DM.
\newblock Learning anisotropic interaction rules from individual trajectories in a heterogeneous cellular population.
\newblock Journal of The Royal Society Interface. 2022;19(195):20220412.

\end{thebibliography}


\end{document}



 \begin{table}
     \centering
     \begin{tabular}{lll}
     \hline
     {\bf Quantity} & {\bf Symbol} & {\bf Dimensions} \\
     \hline

        \hline
        {\em Discrete models} & \\
         Lattice spacing & $\delta$ & $L$ \\
        Time step & $\tau$ & $T$ \\
        Lattice site (in two dimensions) & $(i,j)$ & - \\
        Agent location after $n$ steps & $\vect{X}_n$ & $L$ \\
        $n^\mathrm{th}$ random walk jump & $\vect{Z}_n$ & $L$ \\
        Probability of agent movement per time step & $P_m$ & - \\
        Probability of agent proliferation per time step & $P_p$ & - \\
        Probability of agent death per time step & $P_d$ & - \\
        Bias in movement probabilities & $\vect{\rho}=(\rho_x,\rho_y)$ & - \\
        Spatial domain for agent locations  & $\Omega$ & - \\
        Number of agents in population & $N(t)$ & - \\
        Number of agents at lattice site $(i,j)$ at time $t$ & $U_{ij}(t)$ & - \\
        Expected number of agents at lattice site $(i,j)$ at time $t$ & $u_{ij}(t)= \langle U_{ij}(t) \rangle$ & - \\
        \hline
        {\em Continuous models} & \\        
        Location  & $\vect{x}=(x,y)$ & $L$  \\
        Time  &     $t$ & $T$ \\
        Probability density function for location of a single agent & $p_1(\vect{x},t)$ & $L^{-d}$ \\
        Average density of agents per unit volume  & $p(\vect{x},t)$ & $L^{-d}$ \\
        Average density of agents per lattice site & $u(\vect{x},t)$ & - \\
        Spatial domain for continuum-limit equation & $R$  & -\\
        Diffusivity & $D$ & $L^2T^{-1}$ \\
        Advection velocity & $\vect{v}=(v_x,v_y)$ & $LT^{-1}$\\
        Flux & $\vect{J}$ & $L^{-d+1}T^{-1}$ \\
        Intrinsic (density-independent) net growth rate & $r$ & $T^{-1}$ \\
        Death rate & $\mu$ & $T^{-1}$\\
        Birth rate & $b$ & $T^{-1}$ \\
        Carrying capacity density & $K$ & $L^{-d}$ or -\\
        Displacement vector between agents & $\vect{\xi}$ & $L$ \\
        $k^\mathrm{th}$ spatial moment & $p^{(k)}(\vect{x}_1,\ldots,\vect{x}_k,t)$ & $L^{-kd}$ \\
        Pair correlation function & $g(\vect{\xi})$ & - \\
     \hline
     \end{tabular}
     \caption*{Supplementary Table S1. List of mathematical notation. Bold symbols denote vectors in $\mathbb{R}^d$, where $d$ represents the number of spatial dimensions.   }
\end{table}